**Body-terrain interaction affects large bump traversal of insects and legged robots**

Sean W. Gart and Chen Li

Department of Mechanical Engineering, Johns Hopkins University

3400 N. Charles St, 126 Hackerman Hall, Baltimore, Maryland 21218-2683, USA

E-mail: [redacted]

## Abstract

Small animals and robots must often rapidly traverse large bump-like obstacles when moving through complex 3-D terrains, during which, in addition to leg-ground contact, their body inevitably comes into physical contact with the obstacles. However, we know little about the performance limits of large bump traversal and how body-terrain interaction affects traversal. To address these, we challenged the discoid cockroach and an open-loop six-legged robot to dynamically run into a large bump of varying height to discover the maximal traversal performance, and studied how locomotor modes and traversal performance are affected by body-terrain interaction. Remarkably, during rapid running, both the animal and the robot were capable of dynamically traversing a bump much higher than its hip height (up to 4 times the hip height for the animal and 3 times for the robot, respectively) at traversal speeds typical of running, with decreasing traversal probability with increasing bump height. A stability analysis using a novel locomotion energy landscape model explained why traversal was more likely when the animal or robot approached the bump with a low initial body yaw and a high initial body pitch, and why deflection was more likely otherwise. Inspired by these principles, we demonstrated a novel control strategy of active body pitching that increased the robot's maximal traversable bump height by 75%. Our study is a major step in establishing the framework of locomotion energy landscapes to understand locomotion in complex 3-D terrains.

## 1. Introduction

The ability to traverse large obstacles is crucial for both animals and robots. Many insects [1–5], reptiles [6–9], small birds [10], and small mammals [11] encounter large bump-like obstacles in their natural habitats, such as branch litter and roots on the rainforest floor and rock beds near river or in desert environments (figure 1), which they need to traverse in order to survive [12, 13]. Similarly, the speed at which search-and-rescue robots [14] traverse terrains such as building rubble and landslides where large bumps are abundant (figure 1) could determine the success of failure of a critical mission [15].

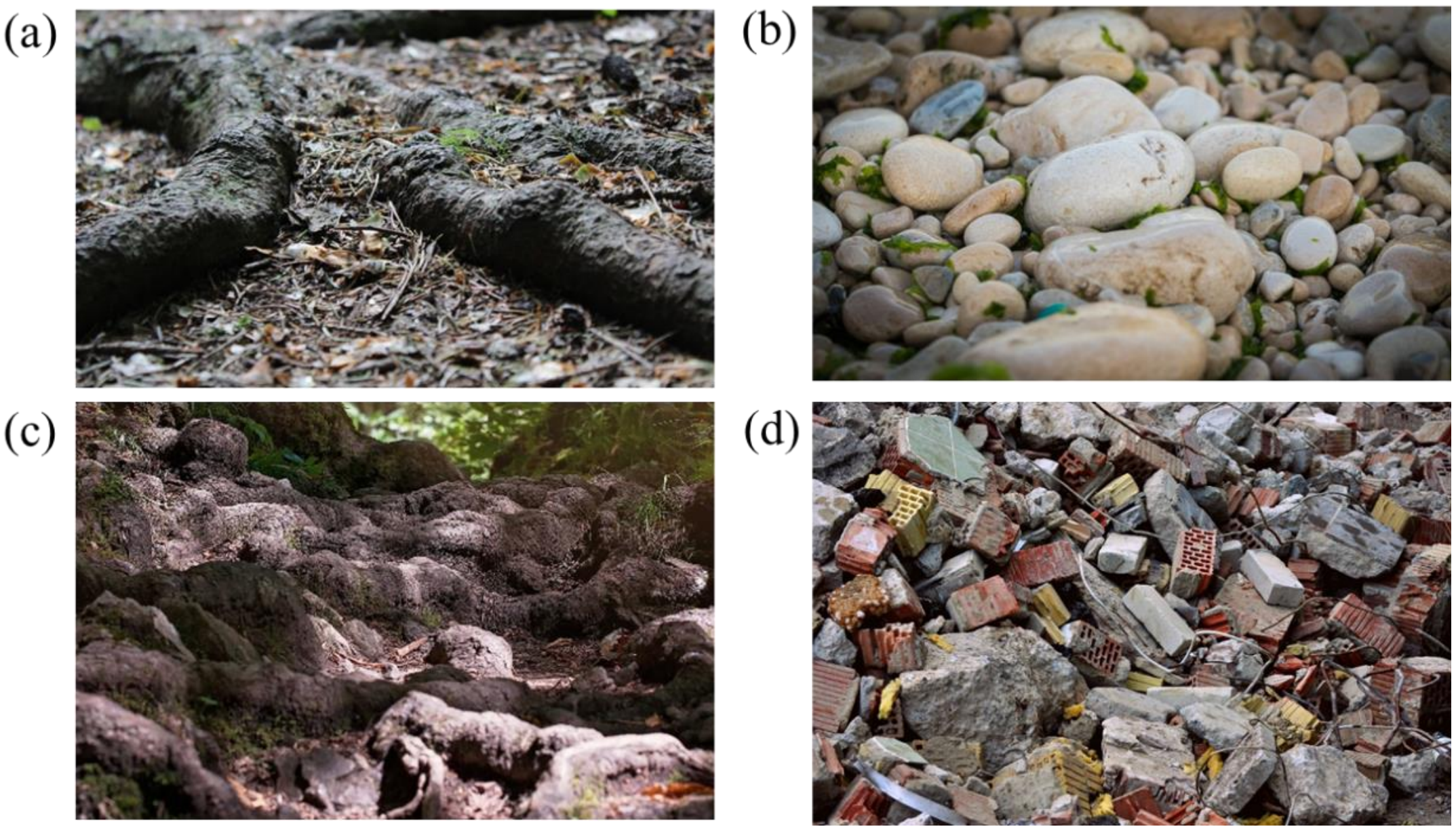

**Figure 1.** Examples of bump-like obstacles in complex 3-D terrains: (a) leaf and branch litter, (b) rock beds, (c) exposed roots, and (d) building rubble. All images were available under Creative Commons CC0 at pixabay.com.

Previous studies of animal locomotion over bump-like obstacles focused on how animals control, adjust, or plan body and leg motions using sensory information [1, 2, 6, 16, 17] to traverse the obstacle, during which body collision with the obstacle is often being actively avoided. Such avoidance of body collision is possible either because the animal starts from at rest or moves slowly enough (lower than 30% of walking-to-running transition speed) [1, 2, 18, 19], and/or because the obstacle is small enough (up to twice the animal's hip height) [8, 20, 21]. Similarly, empirical adjustment of leg control [22] and sensor-based planning [23] have enabled rapid-running legged robots to traverse bump-like obstacles such as stairs, small walls, and blocks, typically up to the robot's hip height [24–28]. Although some robots could traverse a bump-like obstacle beyond one hip height high [24, 29, 30], doing so was realized by ballistic jumping [22, 26, 30] which requires stopping to plan leg movement.

However, during dynamic locomotion in natural and artificial environments, legged animals and robots can sometimes rapidly run into bump-like obstacles even higher than their hip height. In this case, because sensory and neuromuscular delays [31] make it challenging to plan and adjust body and leg movements substantially and rapidly enough [3, 32], avoidance of body collision and physical interaction with obstacles may also be impossible [3, 5, 33–35]. It is not well understood how well animals and robots can traverse bump-like obstacles much larger than their hip height, and whether the often inevitable body-terrain interaction during dynamic locomotion over these obstacles affects traversal.

Unlike locomotion on simpler 2-D surfaces where an animal can walk, run, or climb for extended time, when negotiating large obstacles in such complex 3-D terrains, animals more often use and transition between diverse locomotor modes [6, 19, 35]. Previous studies focused on how animals use sensory information to make decisions about which locomotor mode to use. For example, cockroaches walking at less than 10 cm $s^{-1}$ either climb over, tunnel underneath [19], follow, or turn away from an obstacle [5, 36] depending on how its antennae contacted the obstacle. Recent studies revealed that, during rapid locomotion, the passive physical interaction between animal/robot body with the terrain can also play a major role in initiating diverse locomotor modes while traversing complex 3-D terrains and determining traversal probability, even when no sensory information is available [35]. Our companion paper also demonstrated that passive body-terrain interaction strongly affected traversal performance of large a gap-like obstacle comparable to body size [37].

Inspired by these recent discoveries, we hypothesized that passive body-terrain interaction, if appropriate, can also help animals and robots traverse bump-like obstacles much higher than their hip height. To test this, we observed how well the discoid cockroach and a cockroach-inspired, open-loop legged robot traversed a large bump up to 4 times the hip height, and studied how locomotor modes and traversal performance depended on bump height and body-terrain interaction during bump encounter. Based on these observations, we used a novel modeling framework of locomotion energy landscapes [35] to begin to understand why body-terrain interaction affected locomotor modes and traversal performance. In addition, we compared animal and robot results to gain insights into the role of active sensory feedback. Finally, we demonstrated a control strategy to enhance large bump traversal during dynamic locomotion used our robot with an active tail [37].

## 2. Methods

The majority of methods follow those described in the companion paper [37]. Below we summarize methods used in this study that differed from the companion study.

### 2.1. Experiments

For animal experiments, we used 15 male *Blaberus discoidalis* cockroaches (Pinellas County Reptiles, St Petersburg, FL, USA). The animals measured 4.5 ± 0.8 cm in length and 2.4 ± 0.2 cm in width and weighed 2.7 ± 0.3 g. We used the obstacle track described in our companion study [37], but replaced the gap with a bump that spanned the entire width of the track (figure 2(c)). We tested four bump heights, $H$ = 0.5, 1, 1.5, and 2 cm, ranging from 1 to 4 times the animal's hip height ($z_{hip}$ = 0.5 cm) and collected a total of 553 accepted trials.

To quantify the large variety of locomotor modes during bump traversal, we used an obstacle track capable of high-throughput semi-automated data collection [38]. To automatically stimulate the animal's escape response and encourage them to run over the bump, we attached 3-D printed pushing pads to linear actuators (Progressive Automation, Richmond, BC, CAN) that moved up and down repeatedly to motivate the animal to run from shaded enclaves at each end of the track. The enclaves were shaded from four 500 W work lamps (Coleman Cable, Waukegan, IL, USA) illuminating the middle test section of the track. We constructed a 4 cm long bump with 3-D printed blocks (Ultimaker 2+, Geldermalsen, Netherlands) and covered them with white paper cardstock (Pacon 4-ply railroad poster board, Appleton, WI, USA) to achieve uniform friction properties.

Four synchronized cameras (JAI-5000-PCML, Copenhagen, DEN) filmed the test area from the dorsal view at 50 frames $s^{-1}$ at a resolution of 2560 × 2048 pixels and a 500 μs shutter time. To track the motion of the animal, we attached 10 mm × 10 mm BEEtags [39] dorsally above the body center of mass (CoM). Because tags were occasionally pulled off by pushing pads, we removed the animal's wings before attaching the tags. This did not have an impact on their locomotion because the animal did not contact the bump obstacle with the dorsal surface of its body.

For robot experiments, we tested four bump heights, $H$ = 2.5, 5, 7.5, and 10 cm, ranging from 1 to 4 times the robot's hip height ($z_{hip}$ = 2.5 cm). For each of the four bump heights, we tested four different running speeds (17 ± 7 cm $s^{-1}$, 39 ± 11 cm $s^{-1}$, 61 ± 16 cm $s^{-1}$, and 70 ± 13 cm $s^{-1}$) and three initial body heading angles (0°, 30°, and 60°). We collected 5 trials each for each combination of running speed, initial body heading angle, and bump height, resulting in a total of 240 trials. We constructed the bumps using 30 cm × 60 cm stacked acrylic sheets. To prevent leg slip when the robot started to move, we covered the bottom surface leading to the bump with 50 grit sandpaper. To reduce the friction between the head of the robot and the front face of the bump, we covered the bump surface with polystyrene. For the robot's anterior, we chose to use an ellipsoidal shape similar to the animal head shape to allow direct comparison between animal and robot observations.

For the tailed robot experiment to test our control hypothesis, we tested the tailed robot at a constant speed of $70 \pm 3$ cm $s^{-1}$ with a fixed initial body heading perpendicular to the bump. For both with and without tail actuation, we varied bump height from 1.5 hip height to 4 hip height with an increment of 0.5 hip height. We collected 10 trials for each hip height, resulting in a total of 120 trials.

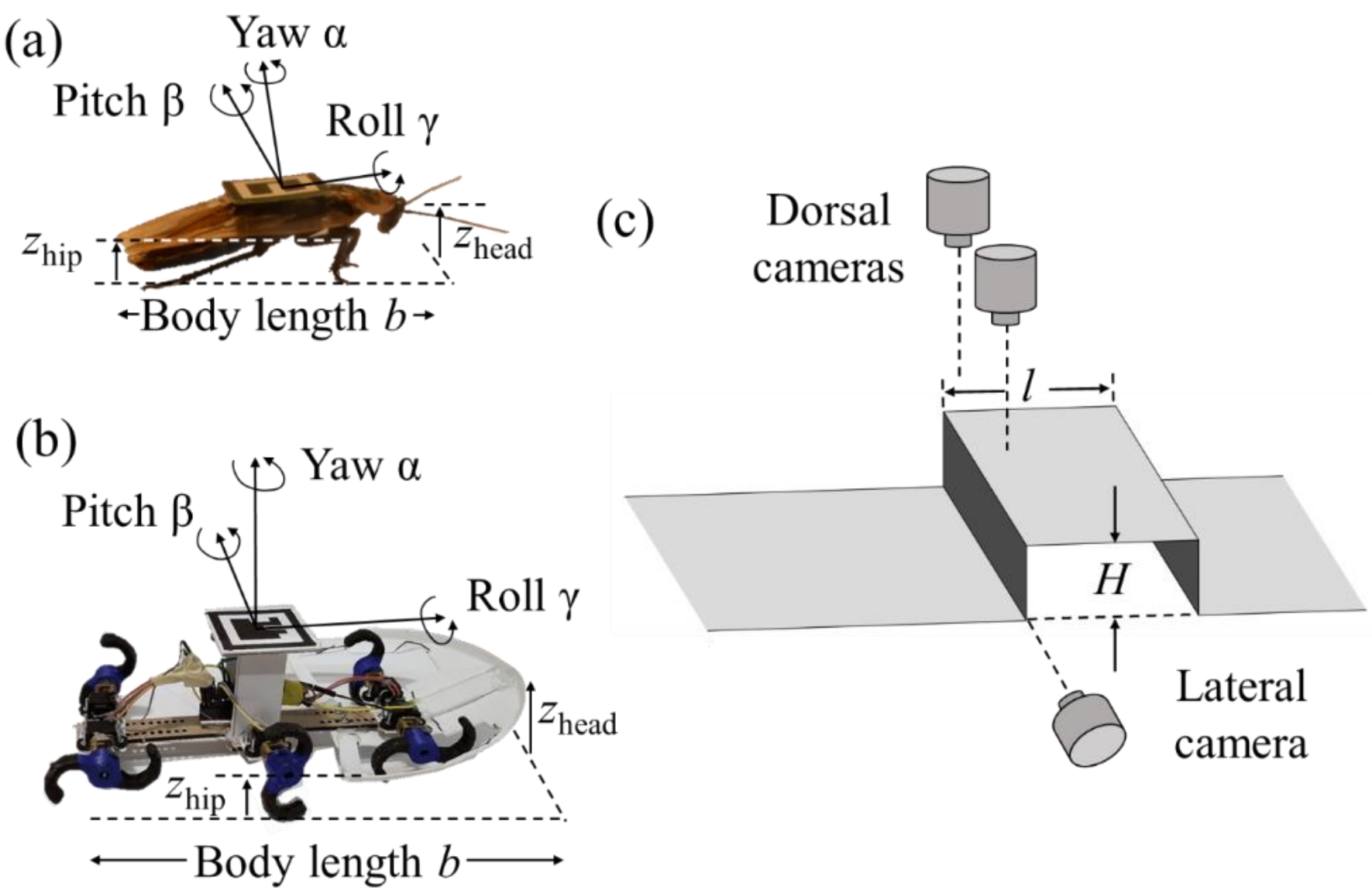

**Figure 2.** Experimental setup and definition of geometric and kinematic variables. (a) Discoid cockroach, hip height $z_{hip} = 0.5$ cm, body length $b = 4.5 \pm 0.8$ cm. (b) Legged robot, hip height $z_{hip} = 2.5$ cm, body length $b = 25$ cm. We used BEEtags [39] to measure 3-D position ($x$, $y$, $z$) and orientation (yaw α, pitch β, and roll γ) of the body. (c) Schematic of the bump obstacle. Bump height $H$ was varied from 1 – 4 hip height and bump length $l = 0.8$ body length for animal experiment and $l = 1$ body length for the robot experiment. Two dorsal cameras and one lateral camera were used to record experiments.

## 2.2. Data analysis

For both animal and robot experiments, we obtained locomotor mode transition ethograms [16, 19, 40] for each traversal attempt. In experiments, we observed four different locomotor modes (see section 3.2): climbing, deflection, simultaneous climbing and deflection, which we referred to as slanted climbing, and flipping over (which only occurred for the robot).

To determine the instantaneous locomotor mode, we measured kinematic data (velocity, body pitch β, and body yaw α) using a moving average over a window of three frames. Transition between modes occurred when two adjacent frames had different locomotor modes. We obtained initial kinematic measurements such as approach speed $v_0$, initial body pitch $\beta_0$, and initial body yaw $\alpha_0$ by averaging data from two video frames prior to bump contact. We observed only a small difference between the velocity heading and body yaw immediately prior to gap encounter (7° ± 8° for the animal, 2° ± 8° for the robot). There we assumed velocity heading always equaled body yaw. Further definitions of these metrics are described in our companion study [37].

To automatically categorize the animal and robot's locomotion during bump negotiation in one of the four modes observed, we quantitatively defined them as follows.

(1) Climbing: For all except the 1 hip height bump, the straight ascent mode occurred when body pitch β exceeded 30° and body yaw was less than 50°. For the 1 hip height bump, because upward body pitch β rarely exceeded 30°, we assumed that if the body was not deflected, the animal or robot traversed in the straight ascent mode.

(2) Deflection: We defined deflection to have occurred when body pitch β was smaller than 30° and either of three criteria were satisfied: body yaw α exceeded 50°; the ratio of lateral velocity to forward velocity exceeded 0.9 at any time after bump collision; or if initial body yaw $\alpha_0$ was greater than 50° and body yaw α increased more than 5° at any time during the bump traversal attempt. These three deflection criteria were necessary because the animal and robot occasionally bounced off of the bump, moving laterally without turning its body yaw α in the lateral direction. Alternatively, it encountered the bump with a high initial body yaw $\alpha_0$ that did not increase during traversal. In this case, the body was moving straight and was not deflected in its heading.

(3) Slanted climbing: The animal and robot climbed with a high body yaw α and velocity heading angle. This mode occurred when body pitch β exceeded 30° and body yaw α exceeded 50° simultaneously. This mode was usually a brief transitional mode between straight ascent and deflection.

(4) Flipping over: Occasionally, the robot body pitched or rolled over 90° and landed on its dorsal side. We only observed this mode in the robot and the animal never flipped over during traversal of bumps up to 4 hip height.

We defined a trial to result in traversal if the center of mass (CoM) crossed the leading edge of the bump.

To check if the animal adjusted its kinematics after antennae contact, we measured approach speed, body pitch, and body yaw when the antennae touched and when the head touched the front face of the bump. We found no statistically significant changes in approach speed and body pitch ($P = 0.113$, $P = 0.995$, respectively, repeated-measures ANOVA) and only a slight increase in body yaw from 21° ± 15° to 25° ± 18° ($P = 0.0002$, repeated-measures ANOVA).

To analyze locomotor pathways for the animal experiment, we first calculated the transition probability between each pair of locomotor modes for all trials from one individual. Then to compare between bump sizes, we averaged the means of all individuals for each bump size. For the robot, we calculated the transition probably between locomotor modes by averaging all trials for each bump height. To further highlight the differences in body orientation between these two locomotor pathways, we averaged head height, body pitch, and body yaw for trials that showed only the straight ascent or only the deflection mode. We excluded diagonal ascent because it was a transitional mode.

To measure bump traversal performance, we measured traversal speed, defined as the average forward speed of the body from when the head reached the near edge of the bump to when the center of mass reached the near edge of the bump.

# 3. Results and Discussion

## 3.1. Animal and robot are capable of traversing a large bump dynamically

Running at an average approach speed of 41 ± 15 cm $s^{-1}$ (independent of bump height, $P = 0.132$, repeated-measures ANOVA), the animal was capable of traversing all the bumps tested, up to the largest 4 hip height bump (figure 3(a), filled circles). Remarkably, the animal was capable of dynamically traversing a large bump at speeds higher than its walking-to-running transition speed with decreasing probability with bump height (figure 3(a), open circles) [41, 42] (traversal speed > 33 cm $s^{-1}$, Froude number $Fr > 1.5$). Additionally, the animal could traverse a bump twice as high as found in previous studies, with a traversal speed up to 4 times higher than previous studies [1, 2] (figure A3, gray areas).

Similarly, the robot was capable of traversing a bump up to 3 hip height (figure 3(b), filled circles), and also often did so dynamically at speeds higher than its walking-to-running transition speed decreasing probability with bump height (figure 3(b), open circles). Despite our robot using open-loop control, its large bump traversal probability was at least 30% higher than achieved in previous studies which robots used carefully planned maneuvers [24–26]. Although the portion of dynamic traversal among all traversal decreased with bump height, it is worth noting that dynamic traversal was possible for all the bump height tested that the animal or the robot was able to traverse (figure 3(c, d)).

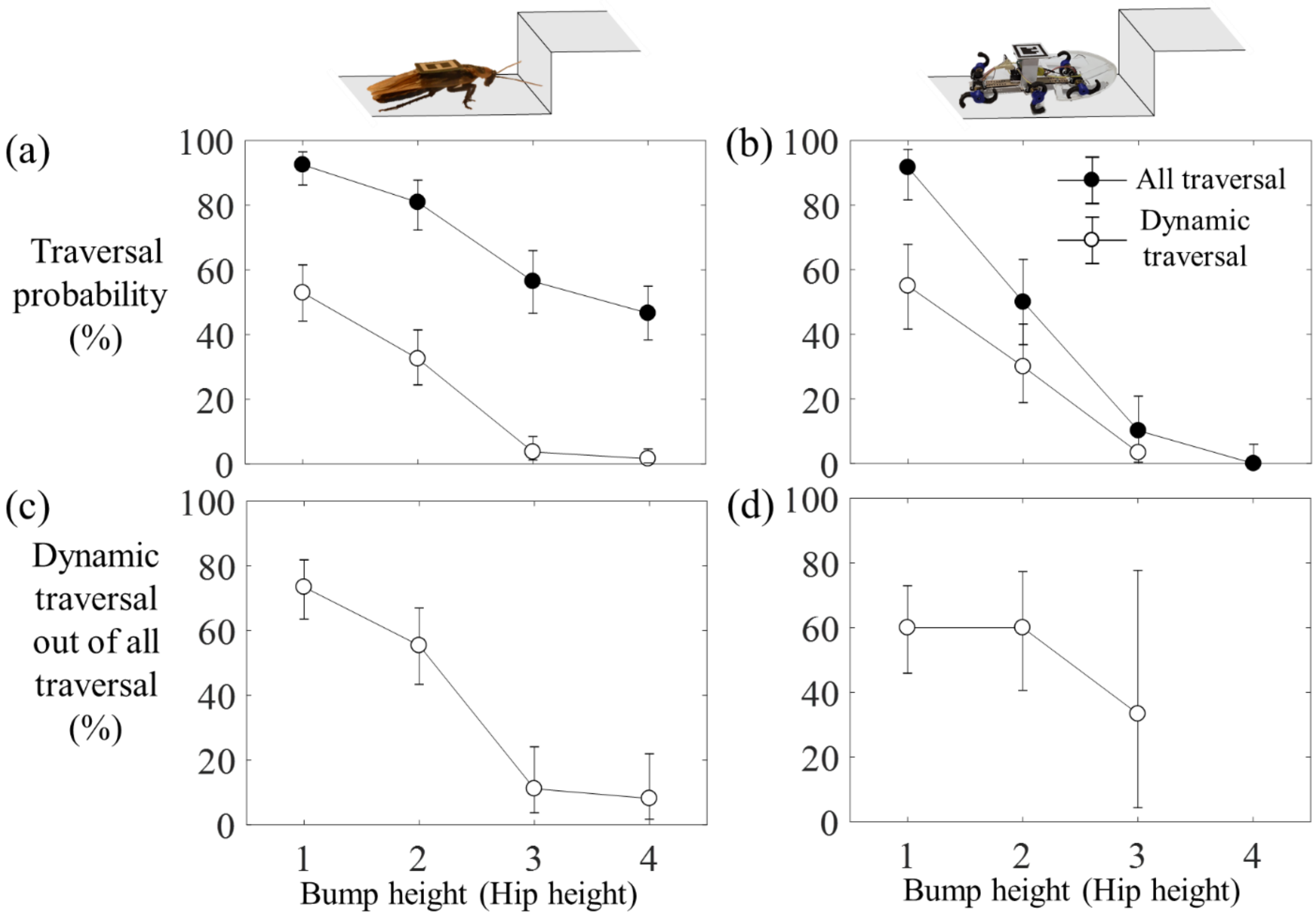

**Figure 3.** Probability of traversal (filled circles) and probability of dynamic traversal out of all trials (open triangles) of the animal (a) and robot (b) as a function of bump height. The animal traversed dynamically if its traversal speed was higher than its walking-to-running transition speed (see methods section 2.2 and the companion paper [37]).

### 3.2. Locomotor modes and transitions

Although the bump obstacle was simple, the animal or robot displayed multiple locomotor modes when attempting to traverse it (see section 2.2 for quantitative definitions of each mode):

(1) Climbing. The animal or robot often climbed against the large bump when its initial body yaw and initial body yaw was small (i.e., body-terrain collision was head-on) (figure 4(a-f), supplementary videos 1, 2). Climbing was most likely to result in successful traversal of the bump. When climbing resulted in traversal, the animal or robot body pitched up to raise the head above the front face of the bump and climbed over it while maintaining nearly 0° body yaw (figures 4(a, f), figures 4(c, d, e, f), red solid curves).

As bump height increased, the animal or robot body pitched up more to traverse (animal: up to 60°; robot: up to 40°), but always recovered the initial orientation within 1 body length of forward displacement once it was on the bump. When the animal climbed a high bump, it gripped the bump with its legs and pulled itself up and forward onto the bump, while flexing the body ventrally up to 23° to allow its hind legs to better brace against the front face of the bump (figure 4(a), frame 4). Although this appeared kinematically similar to previous observations of quasi-static bump traversal [1, 2], in our study, the animal traversed a bump up to twice as high and up to 8 times faster (figure S1, S6).

(2) Deflection. The animal or robot often was deflected against the large bump when its initial body yaw and initial body yaw was large (figure 4(b, g), supplementary videos 1, 2). In this case, the body did not pitch up substantially, nor did the head rise substantially (figure 4(c, h), dashed blue lines). Instead, body yaw increased up to 90°, resulting in turning laterally and failure to traverse (figure 4(e, j), blue dashed curves). Deflection most often resulted in failure to traverse.

It is worth emphasizing that for all but the smallest bump, both the animal and the robot had higher initial head height, higher initial body pitch, and lower initial body yaw when traversing using climbing than when being deflected (figure 4(c, d, e, h, I, j); animal: $P < 0.05$, repeated-measures ANOVA; robot: $P < 0.05$, ANOVA). This is strong evidence that body-terrain interaction played a major role in large bump traversal.

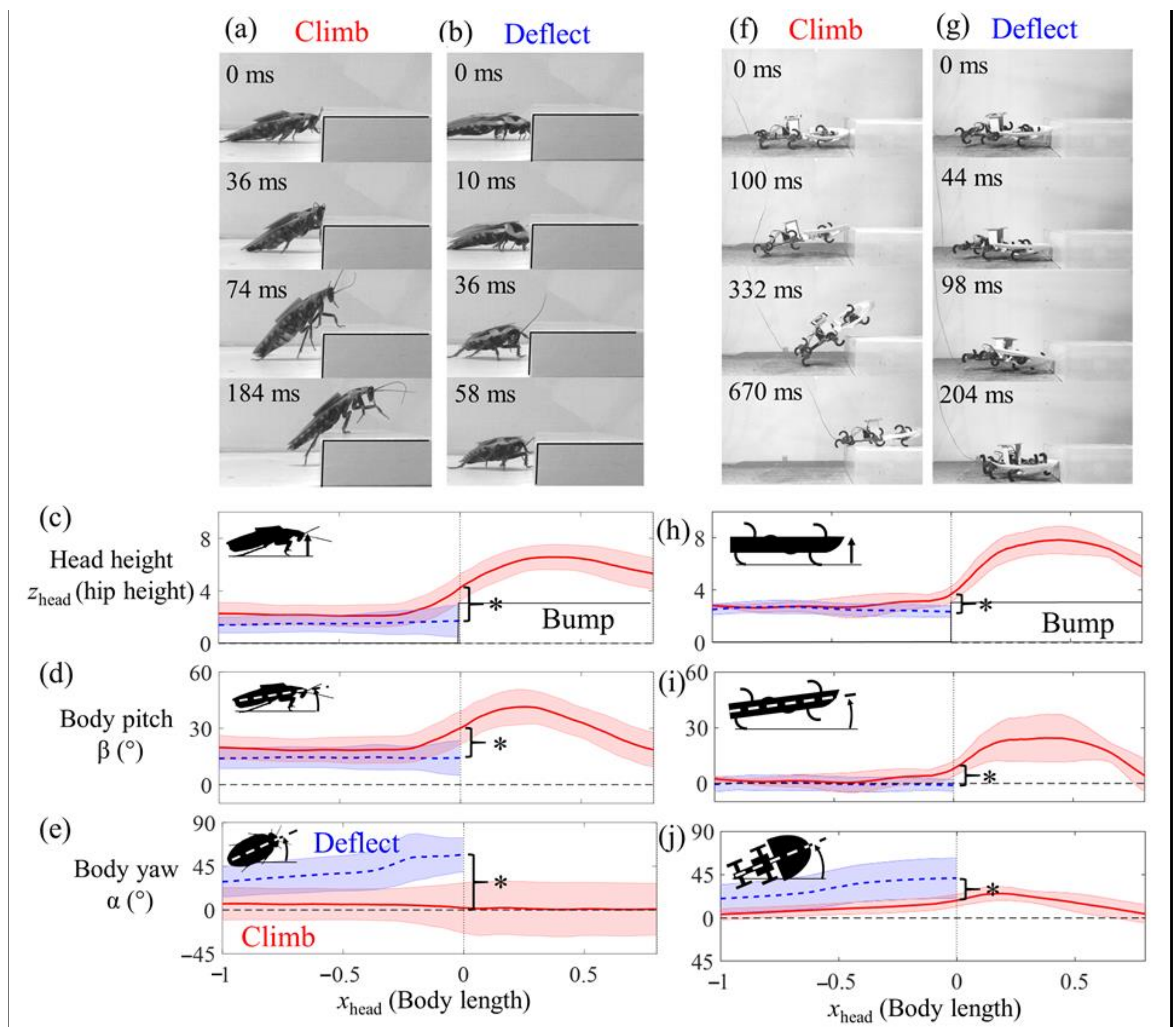

**Figure 4.** Dynamic locomotion of the discoid cockroach and robot attempting to traverse a high bump. Representative trials of (a, f) climbing traversal and (b, g) deflection. (c, h) Head height, $z_{head}$, as a function of *x*-position of the head.* (d, i) Body pitch as a function of *x*-position of the head. (e, j) Body yaw as a function of *x*-position of the head.† Solid red and blue dashed curves and shaded areas represent means ± 1 s.d. for the cases of climbing traversal and deflection. Data are shown for the 3 hip height (animal, 1.5 cm; robot: 7.5 cm) bump as an example and have similar trends for other bump heights. For deflected trials,

* We noted that the head appeared to have slightly penetrated the bump after collision. This was due to the robot head and body flexion after high-speed collision with the bump, which our head tracking method could not account for (see methods in our companion paper for details [37]).

† Due to the ellipsoid-like shape of the animal body and the robot anterior, the head was only able to reach the bump ($x_{head}$ = 0) for body yaw α < 45. As body yaw continued to increase during deflection, the head forward position $x_{head}$ actually decreased).

data were only shown until bump collision because the animal often moved backwards afterwards. Bracket and asterisk represent statistically significant differences.

(3) Slanted climbing. Occasionally, the animal and robot climbed even with a high initial body yaw and high initial body yaw, resulting in a slanted trajectory relative to the bump (figure 5(a), dark yellow rectangle) as opposed to the trajectory perpendicular to the bump using head-on climbing. Careful observation revealed that, during slanted climbing, the animal often transitioned between climbing and deflection, with body yaw and velocity heading both oscillating by larger amplitudes that during head-on climbing. This occurred more frequently as bump height increased ($P < 0.0001$, repeated-measures ANOVA). To transition from the deflection to climbing, the animal often gripped and hung on to the front face of the bump with a fore leg, pitching its body upwards and turning towards the bump with decreasing body yaw. Once the second fore leg was able to reach the top of the bump, the leg and body kinematics were similar to that during climbing. Because the robot operated with feed-forward control, its slanted climbing was merely due to oscillations in body yaw and body pitch from intermittent leg-terrain interaction. The robot did not actively transition between climbing and deflection like the animal did.

(4) Flipping over. The open-loop robot occasionally flipped over onto its dorsal side and failed to traverse after impacting a large bump obstacle (3 and 4 hip height, figure 5(c, d), purple oval). The animal was never flipped over in our experiments.

In addition, we found that the animal did not always use one locomotor mode when attempting to traverse the large bump but frequently transitioned between modes, forming complex locomotor transition pathways (figures 5). Unlike many previous studies [1, 3, 43, 44] where the focus was on a single locomotor mode during obstacle traversal, such comprehensive observations of locomotor transition pathways [6, 18, 38, 45] provided a more realistic representation of animal locomotion in nature.

Comparison of locomotor transition pathways (figure 5) between each bump height revealed that, as bump height increased, the animal or robot less frequently climbed and traversed, but were more often deflected (animal: $P < 0.0001$; robot: $P < 0.0001$, multiple logistic regression). This change in the dominant locomotor mode was the main reason why the traversal probability decreased with bump height.

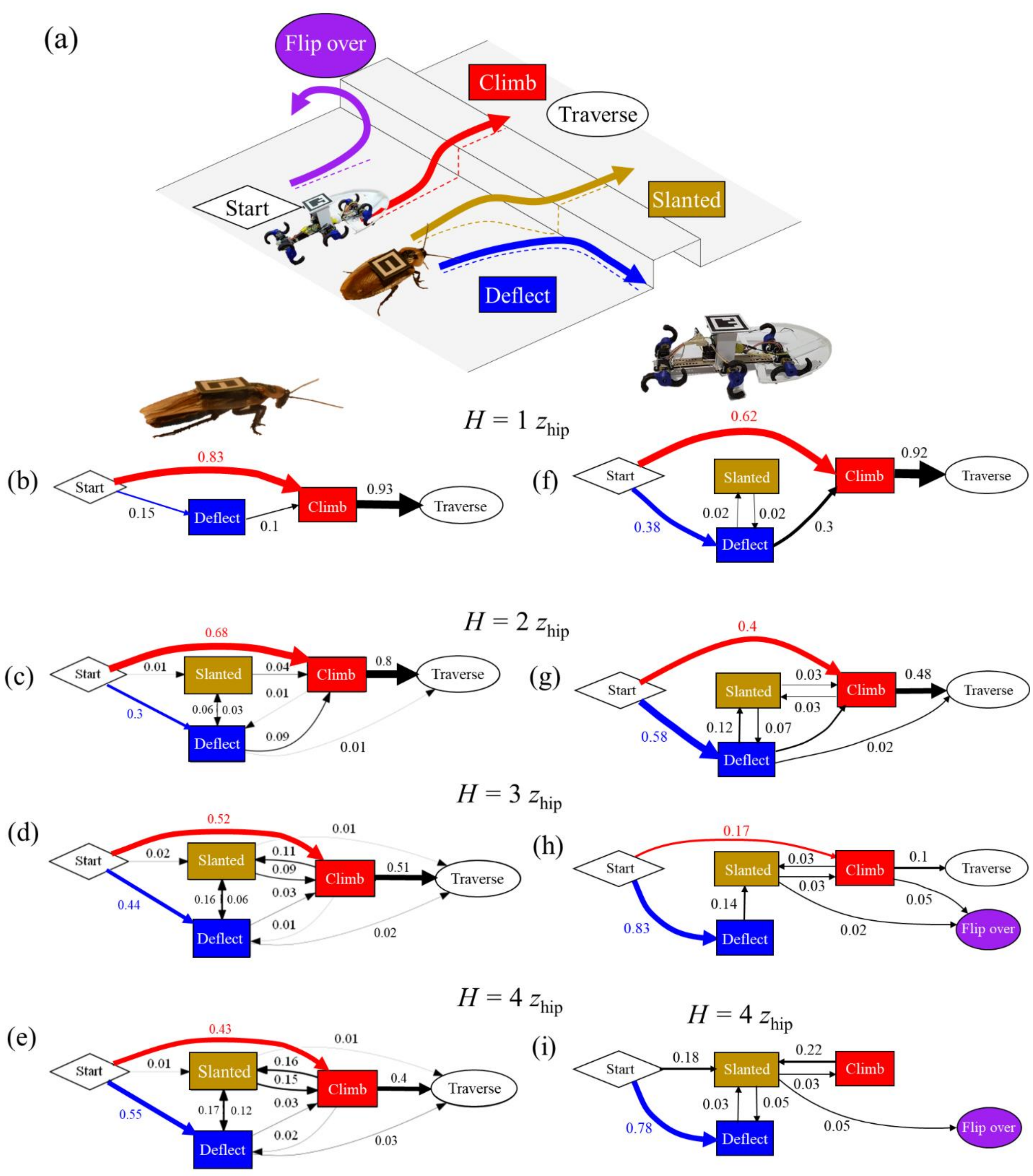

**Figure 5.** The animal and the robot used a variety of locomotor modes while attempting to traverse a large bump. (a) Representative drawings of each locomotor mode. All trials begin with start (white diamond), use one or a sequence of three modes—climbing (red rectangle), deflection (blue rectangle), slanted climbing (dark yellow rectangle), and end with traversal (white oval), flip over (purple oval), or failure to traverse (exiting the field of view, not shown). The arrows represent the typical trajectory of the head for each locomotor mode and the dashed curves show the projection of the trajectory to the surface. Animal (b, c, d, e) and robot (f, g, h, i) locomotor pathways for each bump height (1, 2, 3, and 4 hip height, respectively). Arrows indicate transition between modes. Arrow thickness is proportional to transition probability, indicated by the number near each arrow.

### 3.3. Body-bump interaction is frequent

We observed that, besides legs, the animal's body frequently collided and continued to physically interact with the bump during traversal, especially for larger bumps (figure 6). For the 2, 3, and 4 hip height bump, collision probability was 34%, 61%, and 83%, respectively (figure 6, filled circles). This is in contrast to previous studies on obstacle avoidance [5, 16, 17] or step climbing [1, 2, 6] where movement was slow and obstacle collision was rare. The high probability of body-bump collision was because the animal ran so fast that it did not have enough time to slow down even if antennae touched the bump. For the majority of our trials (83%), the animal ran at speeds higher than the reaction time limit (figure A1, see section 3.2 in our companion study [37]). Similarly, the open-loop robot also collided with the bump (figure 6, open circles) and continued to physically interacted with it frequently during traversal.

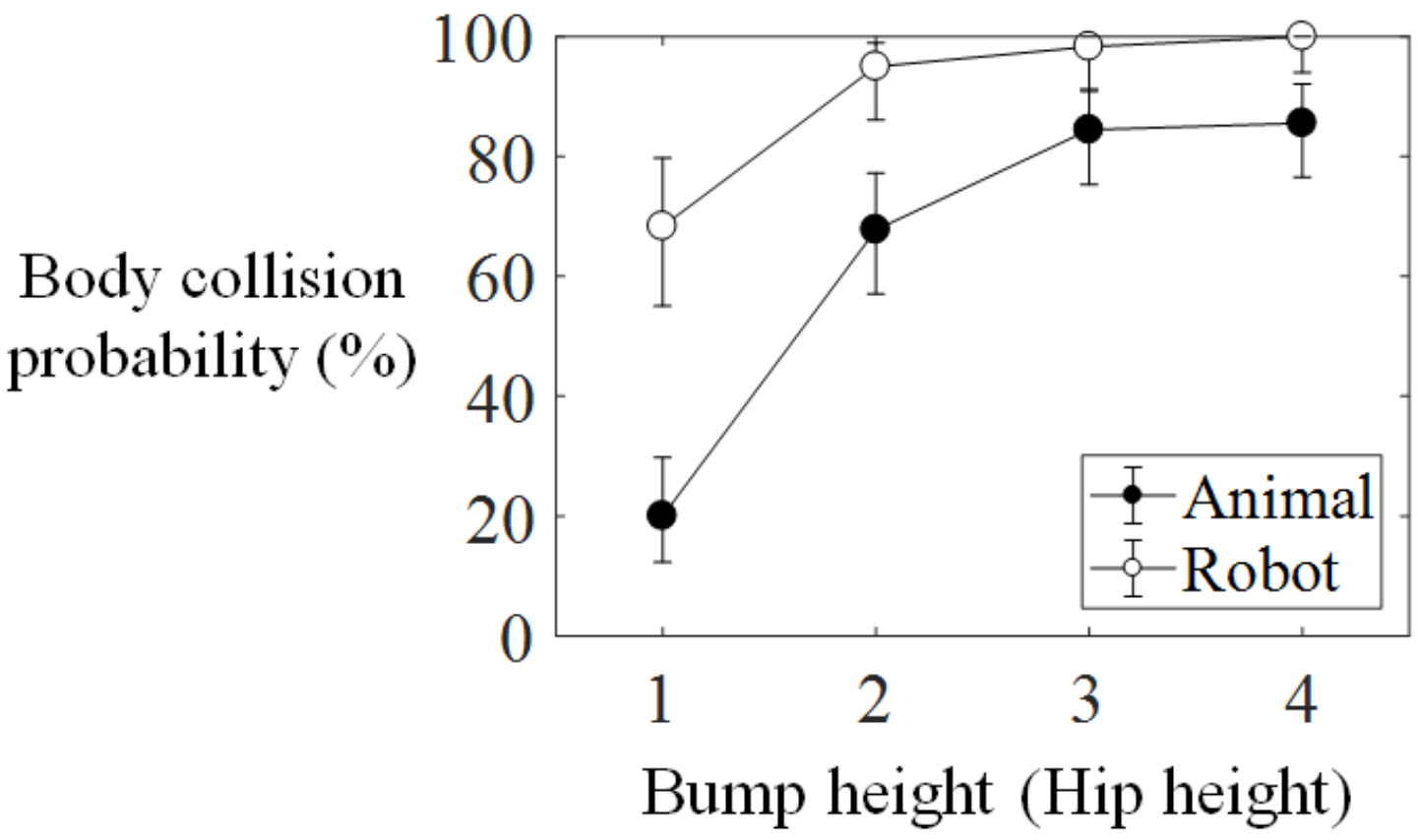

**Figure 6.** The animal and the robot frequently collided with a large bump. Animal (filled circles) and robot (open circles) probability of the body colliding with a bump increased with bump height. Error bars show 95% confidence interval.

It is worth emphasizing that these observations differed from previous studies of cockroach bump traversal, where locomotion was much slower (up to 8 times) and the animal had sufficient time to use its antenna to detect the bump, then slowed down or stopped and adjusted leg and body kinematics to traverse or turn laterally, without body-bump collision [1, 19, 46].

### 3.4. Locomotion energy landscape model

Considering how frequent the animal or robot's body collided with and continued to physically interact with the bump in our experiments, we speculated that body-bump interaction played an important role in the observed sensitive dependence of traversal performance and locomotor transition pathways on bump height. To understand this, we developed a simple locomotion energy landscape model [35] (figures 8, 9, supplementary video 3). The new concept of locomotion energy landscapes was recently introduced to model body-terrain interaction during traversal of complex 3-D terrains such as grass-like beam obstacles, where comprehensive contact mechanics models are still lacking [35]. Although locomotion energy landscapes do not yet model leg-terrain contact or take into account the dynamics and non-conservative forces of the system, they are useful in explaining how body-terrain interaction affect locomotor transition pathways and traversal performance [35].

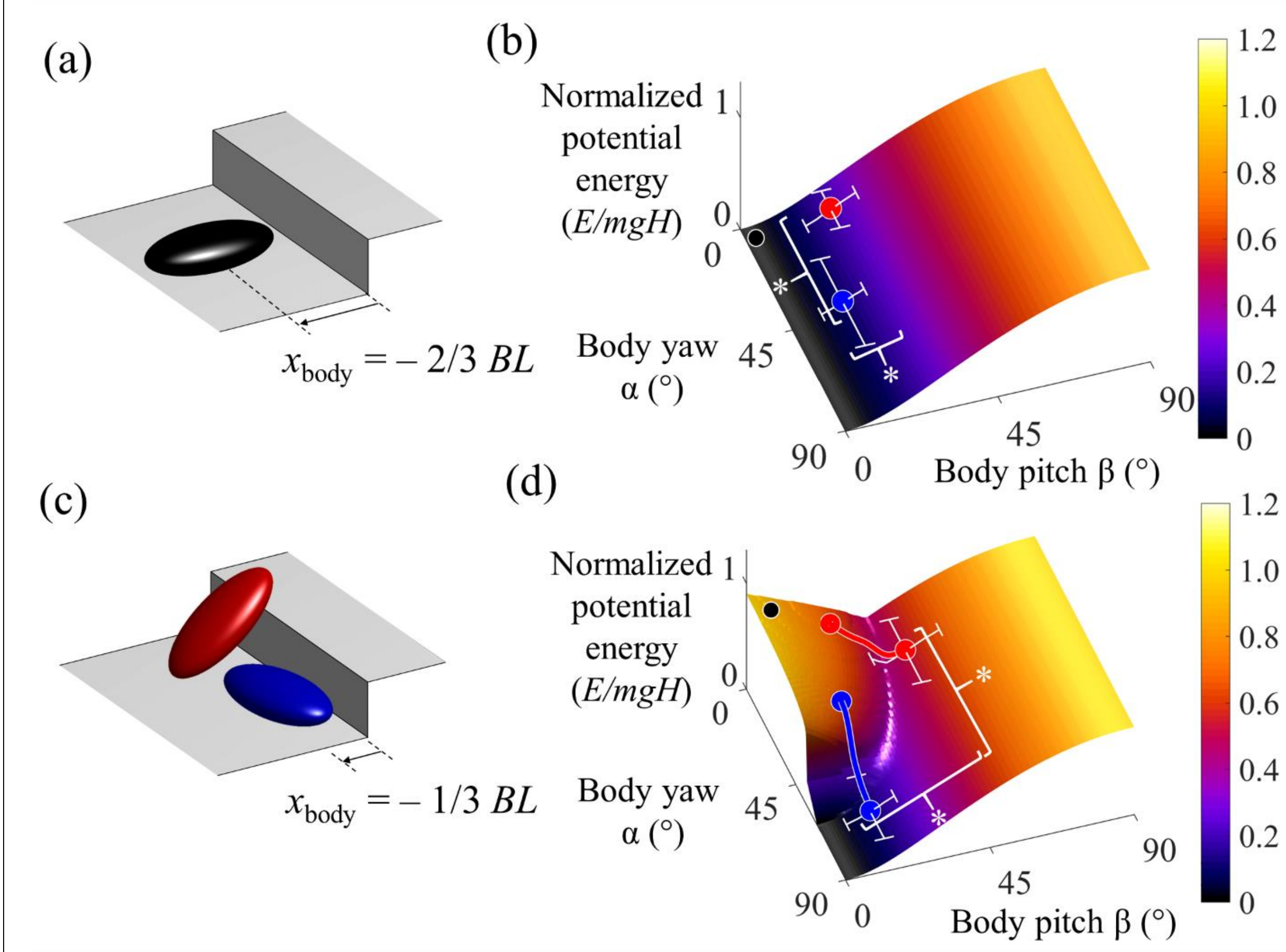

**Figure 7.** Locomotion energy landscape model for 4 hip height bump traversal. (a, c, e) Body forward position relative to the bump and body orientation. The ellipsoidal bodies show the body orientation prior to bump encounter (black) and for climbing (red) and deflection (blue). (b, d, f) Locomotion energy landscape. The energy landscape surface is a function of body yaw, body pitch, and $x$-position perpendicular to the bump and is normalized by the body mass times bump height ($E/(mgH)$). The model is shown at two representative locations: (a, b) just prior to the body colliding with the bump, $x_{body} = -2/3$ body length, (c, d) during traversal, $x_{body} = -1/3$ body length. Red and blue circles indicate the animal's mean initial body pitch and mean initial body yaw during pure climbing and pure deflection, respectively. Error bars represent ± s.d. Red and blue trajectories show the evolution of mean body pitch and mean body yaw as the animal moved forward from $x_{body} = -2/3$ to $x_{body} = -1/6$ body length during pure climbing and pure deflection, respectively. Brackets and asterisk represent a statistically significant difference.

We approximated the animal or robot body as a rigid ellipsoid with the same length, width, and thickness as measured in experiments and a uniform mass distribution such that center of mass (CoM) overlapped with geometric center. We assumed that the body had no overlap with the bump during traversal and that its lowest point always touched the ground before encountering and after traversing the bump. In addition, for simplicity, we approximated body roll as constantly zero because the animal or robot's body roll oscillated around 0° by only a small magnitude (± 4°) and did not have a significant impact on traversal (animal: $P > 0.05$, repeated-measures ANOVA; robot: $P > 0.05$, ANOVA). These constraints, together with the invariance of the system in the lateral ($y$) direction, meant that the center of mass height, and thus the gravitational potential energy of the body, depended only on its forward position relative to the bump, body pitch, and body yaw, i.e., $E = mgz_{\mathrm{CoM}}$, and $z_{\mathrm{CoM}} = z_{\mathrm{CoM}}(x, \alpha, \beta)$.

To obtain the potential energy landscape resulting from body-bump interaction, we varied the forward ($+x_{\mathrm{body}}$) position of the body, and numerically calculated how potential energy depended on body yaw, body pitch, and body roll (supplementary video 3). To understand how the locomotion energy landscape changed as the animal or robot moved forward towards the bump during traversal (with increasing $x_{\mathrm{body}}$), we examined the landscape at two representative $x$ positions (figure 7). For the following discussions, we used results of a 4 hip height bump as an example. Modeling results of other bump heights were qualitatively similar.

When the body was far away from the bump ($x_{\mathrm{body}} < -1/2$ body length, figure 7(a, b)), its potential energy only depended on body pitch (assuming constant zero body roll, see above). As the body moved closer to the bump ($-1/2$ body length $< x_{\mathrm{body}} < 0$ body length, figure 7(c, d)), it had to turn laterally, pitch up, and/or rise so as not to penetrate the bump. With no change or only a small change in body pitch or body yaw, the body must rise over the edge of the bump so as not to penetrate it. This resulted in a large "bump" in the potential energy landscape for small body pitch and body yaw (figure 7(d), black circle). As body pitch and/or body yaw increased, the amount that the body had to rise decreased. This resulted in a monotonic decrease of the potential energy as body pitch and body yaw initially increased from zero (e.g., figure 7(d), red and blue curves). For a given body yaw, as the body pitch continued to increase, the potential energy reached a local minimum and then increased, resulting in a "valley" in the energy landscape. This valley corresponded with all the body orientation states where the body was in contact with the ground and the top edge or front face of the bump at the same time. The valley became deeper as body yaw increased and eventually reached a potential energy well (global minimum) (figure S2).

### 3.5. Model explained dependence of bump traversal on body-terrain interaction

A quasi-static stability analysis using the topology of the potential energy landscape provided insight into how initial body pitch and initial body yaw affected locomotor modes for a given bump height. The potential energy landscape during body-bump contact had a saddle point, i.e., the local minimum in the body pitch direction at a finite positive body pitch and zero body yaw (figure 7(d)). As the body moved closer and came into contact with the bump, a pitched-up body was most stable along the pitch direction (stable equilibrium in the pitch direction), but this pitched-up body orientation was unstable in the yaw direction (unstable equilibrium). This suggested that, except when the animal or robot always approached the bump head-on (maintaining zero body yaw), it tended to be deflected to either side (increasing body yaw) as long as there was any slight body yaw oscillation (figure 7(d), blue curve). In addition, when the body was deflected (increasing body yaw), it became attracted towards the potential energy well along the valley where it was eventually trapped in an orientation with high body yaw and low body pitch (bottom of figure 7(d)). During slanted climbing, the animal had to overcome this tendency to be deflected further. Finally, the model also provided insight into how bump height affected body-terrain interaction and thus traversal performance. As bump height increased, we found that the potential energy well became deeper and more difficult to escape (figure S2).

These stability analyses revealed two general principles for large bump traversal during which body-terrain interaction played a major role. First, a head-on approach (small initial body yaw) and a pitched up body posture facilitate traversal using climbing. Indeed, we observed that for all except the smallest bump height, both the animal and the robot had a low initial body yaw and a high initial body pitch when they traversed the bump via climbing (figure 4(c, d, e, h, i, j), figure S4(b, c, d, e)). By contrast, when they were deflected and failed to traverse, initial body yaw was significantly higher (animal: $P < 0.0001$; robot: $P < 0.0001$, multiple logistic regression). Second, as bump height increases, traversal via climbing becomes less likely, because it is more difficult to escape the deeper potential energy well that the body can get deflected into. Indeed, we observed both the animal and robot were more often deflected with lower traversal probability as bump height increased (animal: $P < 0.0001$; robot: $P < 0.0001$, multiple logistic regression).

Finally, the model also suggested that a pitched-up body posture, as opposed to a horizontal posture (zero body pitch), also facilitated climbing and traversal. Examination of the potential energy landscape as a function of body pitch and body forward position (figure S3) showed that, when initial body pitch was near zero, the body had to climb up a steep, high potential barrier over a very small forward displacement. Just a slight increase of body pitch reduced the slope of this potential energy barrier, making traversal easier.

We noted that all these analyses were using a simple quasi-static body-terrain interaction model and neglected body and leg dynamics. In reality, the animal or robot ran into the bump with high speed and pushed its legs against the ground during traversal. Ground reaction forces, together with high body momentum (if properly re-directed), could also played a role in traversal of large bump obstacles during dynamic locomotion, and should be better understood in future work.

**3.6. Tail-assisted pitch control to enhance high bump traversal**

The experimental observations that successful traversal had significantly higher initial body pitch (figure 4(d, i), figure 7(b)) and higher initial head height (figure 4(c, h)) and the above model insights that increasing body pitch facilitated climbing inspired us to propose a novel control strategy to use active body pitching to increase dynamic bump traversal performance of the robot. To demonstrate this, we tested the tailed robot developed in our companion study [37].

While running at a constant running speed of $70 \pm 3$ cm $s^{-1}$, the robot's active tail increased its initial body pitch from $0° \pm 4°$ to $3° \pm 7°$ ($P = 0.04$, ANOVA). This small increase in initial body pitch not only increased traversal probability for all except the smallest bump tested, but also increased the maximal traversable bump height for the tailed robot from 2 hip height to 3.5 hip height, a 75% increase (figure 8).

Although the increase in initial body pitch by the active tail appeared small, it raised the head by 1.3 cm, about 15% of the height of the highest bump traversable by the robot (3.5 hip height). This likely increased the probability of the robot's head to reach over the front corner of the bump (figure S3).

We noted that the timing of tail actuation was important for pitching up the body at the right moment so that the head reached over the top of the bump, particularly for the larger 3 and 4 hip height bumps. Pitching up too late did not allow the head to clear the bump in time, while pitching up too early did not work because the increase of body pitch using an active tail was temporary. Future studies should add fast sensors to detect impending bump obstacles during rapid locomotion and precisely control the magnitude and timing of active body pitching [47–49] to further increase traversal performance.

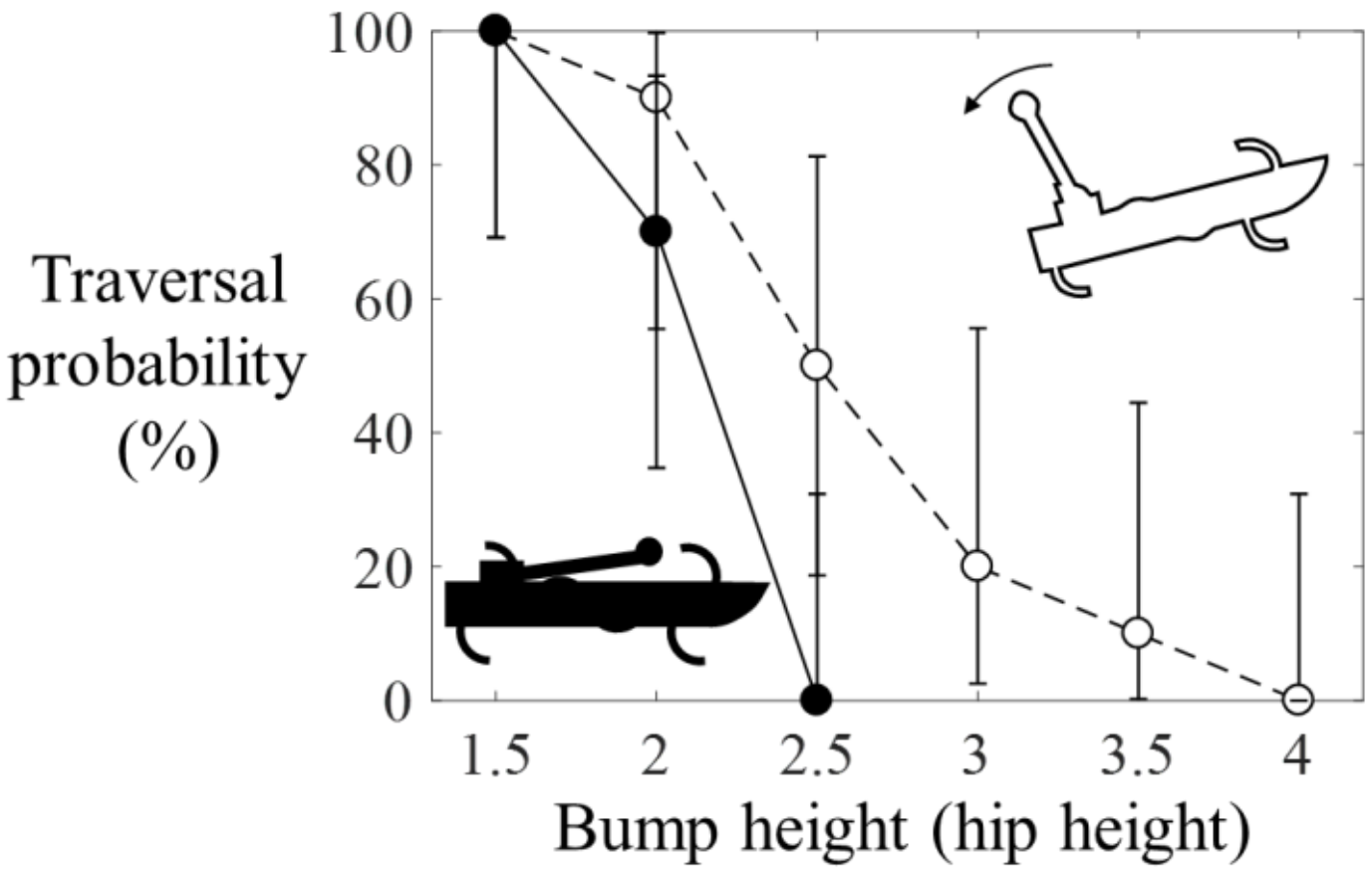

**Figure 8.** Bump traversal probability for the tailed robot, with (open circles) and without (filled circles) tail actuation as a function of bump height.

### 3.7. The role of sensory feedback in traversal

Although the rapid-running animal frequently collided with the high bump and body-terrain interaction strongly affected whether it climbed to traverse or was deflected, the animal likely used sensory feedback to actively adjusted its body and leg kinematics for traversal, as well as to overcome less advantageous body-bump interaction. This was evidenced by a few observations. First, during climbing, especially for larger bump height, the animal often actively flexed its neck joint and abdomen and used its legs to grip and pull itself onto the bump. Second, during slanted climbing, although the animal was susceptible to being deflected given its large initial body yaw, it only occasionally lost grip and eventually become deflected (6% and 12% for the 3 and 4 hip height bump, respectively, figure 5(d, e, h, i)). Third, the animal was still able to traverse a 3 hip height bump even when initial body yaw was up to 60° for the 3 hip height bump and up to 50° for the 4 hip height bump.

By contrast, the sensor-less, open-loop robot had a poorer ability to grip with its legs and lacked body flexibility and had no ability to actively adjust its body orientation or leg kinematics. Compared to the animal, it more often failed by falling backward during climbing (figure 5(h, i)). In addition, the robot was unable to traverse a 3 hip height bump when initial body yaw exceeded 30° and was never able to traverse the 4 hip height bump. Further, the robot occasionally lost stability during climbing and flipped over, which was never seen in the animal experiment. All these limitations led the robot to have worse traversal performance than animal, particularly for the large bump height figure 5(e, i). We expect that adding controlled body flexion guided by the locomotion energy landscape model could further increase maximal traversable bump height for robots—a previous study had doubled this limit using empirically tuned body flexion [23, 29, 50]. Future studies should also add adhesive mechanisms [51–53] to improve the robot's gripping ability.

Finally, previous studies have shown that whether cockroaches can locate the top of a bump with its antennae determines whether it climbs over or under a shelf during slow locomotion [19]. When the animal's body was more pitched up with its head higher and redirected upwards during collision, it might have been better able to locate the top of the bump with its antennae and further make kinematic adjustments to climb [1, 2]. By contrast, when the animal's initial body pitch was low and its head deflected laterally, it might not have been able to do so and instead perceived the bump as a wall-like obstacle and continued to turn and follow the wall [54, 55].

### 3.8. Similarity between large bump and gap traversal

Comparing the current study with our companion study [37], we found that, surprisingly, dynamic traversal of large gap and bump obstacles share a few common features. First, it was possible to dynamically traverse both a large gap and a high bump comparable to body size (and much larger than observed in previous studies [1, 2, 19, 42]), even for the open-loop robot. In addition, during rapid dynamic locomotion over such a large obstacle, because body-terrain contact was nearly un-avoidable and sensor-based adjustment and planning became unfeasible, body-terrain interaction had a strong impact on traversal. Despite the distinct differences between a large gap and a high bump, traversal of both obstacles was facilitated by rapid running towards the obstacle with a pitched-up, head-on body orientation. Finally, for both obstacles, the animal out-performed the robot because it was able to better grip and brace itself using sensory feedback, and active body pitching increased the robot's maximal traversal performance.

## 4. Conclusions

In this study, we discovered that small insects and legged robots are capable of dynamically traversing a bump obstacle much higher than its hip height. Remarkably, the discoid cockroach was capable of dynamically traversing a large bump up to 4 times its hip height at speeds comparable to [41, 56] or higher [1, 2, 57–61] than during running, and our open-loop robot was capable of dynamically traversing a high bump up to 3 times its hip height at speeds above the walking-to-running transition speed. These newly discovered performance limits for dynamic bump traversal were more than 200% that of previous animal studies [1, 2] and more than 30% higher than previous robot studies [24–26]. In addition, we discovered that body-terrain interaction strongly affected locomotor modes and traversal performance. Furthermore, our locomotion energy landscape model explained why traversal was less likely as bump height increased and why a head-on and slightly pitched-up body orientation facilitated dynamic bump traversal by reducing the likelihood of being deflected. These experimental and modeling insights allowed us to use active body pitching to increase maximal traversable bump height by 75%. Finally, the animal's traversal performance exceeded that of the robot thanks to its ability to use sensory feedback to make body and leg adjustments and its better gripping ability and body flexibility.

This study also expanded the novel approach of locomotion energy landscapes [35] in modeling body-terrain interaction to large bump obstacles, another representative 3-D terrain. Besides sensor-based planning and decision processes using geometric information of the environment [19, 46, 62, 63], physical interaction of animal/robot body with the terrain also played an important role in initiating locomotor transitions and determining performance during large obstacle traversal. Together with recent success of this approach in understanding how locomotor shape affect traversal of cluttered grass-like obstacles [35], our study is further establishing locomotion energy landscapes as a general framework for modeling locomotor-terrain interaction in a diversity of complex 3-D terrains.

Together, our two companion studies ([37] and this paper) are a major step towards establishing the new field of terradynamics of animal and robot locomotion in complex terrains. First, the remarkable performance limits of dynamic traversal of large gap and bump obstacles from our studies provided new insights into how well and by what means animals can move in the complex natural world. The ability to dynamically traverse large gaps and bumps comparable to body size during rapid predator-prey pursuit is crucial for the survival of insects and many other legged animals that live on forest floor [64], in scrubs [9] and crop fields [65, 66], and in rocky environments [13]. Our studies provide the first mechanistic explanations of how the need to survive exerts selective pressure on the ability of both predators and prey to run fast and maintain appropriate body posture so that traversal is likely and without significant stumble or slow down, even when sensing becomes unfeasible.

Second, we expect that the performance limits, mechanical principles, and novel control strategies revealed by our two companion studies will inspire a variety of robots to better take advantage of kinetic energy and momentum of rapid running and simple body posture control to traverse large gap- and bump-like obstacles. This will not only add high-speed dynamical traversal of large gap and high bump obstacles to the locomotor repertoire of legged robots, but also simplify their lower-level control so that sensing and computation resources can be better devoted to deliberate maneuvers and precise planning [24–26, 30] required for truly large un-traversable obstacles beyond the limits by using dynamics. Such advancements will expand the operation time and accessible terrain area for robots that perform tasks in complex 3-D terrains such as building rubble and landslides.

Finally, considering their distinct differences in geometry, it is surprising that similar principles exist for traversal of both a large bump and a large gap during rapid locomotion. This suggests that there may be general principles of dynamic locomotion in a diversity of complex 3-D terrains. Future terradynamic studies should further explore the locomotor and terrain parameter space to reveal these general principles and enable quantitative understanding and predictions of the movement of terrestrial animals and robots.

## Acknowledgements

We thank Yuanfeng Han, Tom Libby, Simon Sponberg, Bob Full, and two anonymous reviewers for helpful discussions; Nastasia Winey and Rafael de la Tijera Obert for help with preliminary experiments; and Yuanfeng Han for help with experimental setup. This work is funded by a Burroughs Wellcome Fund Career Award at the Scientific Interface, an Army Research Office Young Investigator Award, and The Johns Hopkins University Whiting School of Engineering start-up funds to C L. Author contributions: S W G designed study, performed experiments, analyzed data, developed model, and wrote the paper; C L designed and supervised study, assisted with model development, and wrote the paper.

## References

[1] Watson J T, Ritzmann R E, Zill S N and Pollack A J 2002 Control of obstacle climbing in the cockroach, Blaberus discoidalis. I. Kinematics *J. Comp. Physiol. A* **188** 39–53

[2] Watson J T, Ritzmann R E and Pollack A J 2002 Control of climbing behavior in the cockroach, Blaberus discoidalis. II. Motor activities associated with joint movement *J. Comp. Physiol. A* **188** 55–69

[3] Zurek D B and Gilbert C 2014 Static antennae act as locomotory guides that compensate for visual motion blur in a diurnal , keen-eyed predator *Proc. R. Soc. B* **281** 20133072

[4] Lewinger W a., Harley C M, Watson M S, Branicky M S, Ritzmann R E and Quinn R D 2009 Animal-inspired sensing for autonomously climbing or avoiding obstacles *Appl. Bionics Biomech.* **6** 43–61

[5] Baba Y, Tsukada A and Comer C M 2010 Collision avoidance by running insects: antennal guidance in cockroaches. *J. Exp. Biol.* **213** 2294–302

[6] Kohlsdorf T and Biewener A A 2006 Negotiating obstacles: Running kinematics of the lizard Sceloporus malachiticus *J. Zool.* **270** 359–71

[7] Tucker D B and Mcbrayer L D 2012 Overcoming obstacles: The effect of obstacles on locomotor performance and behaviour *Biol. J. Linn. Soc.* **107** 813–23

[8] Olberding J P, McBrayer L D and Higham T E 2012 Performance and three-dimensional kinematics of bipedal lizards during obstacle negotiation *J. Exp. Biol.* **215** 247–55

[9] Parker S E and McBrayer L D 2016 The effects of multiple obstacles on the locomotor behavior and performance of a terrestrial lizard *J. Exp. Biol.* **219** 1004–13

[10] Birn-Jeffery A V, Hubicki C M, Blum Y, Renjewski D, Hurst J W and Daley M A 2014 Don't break a leg: running birds from quail to ostrich prioritise leg safety and economy on uneven terrain. *J. Exp. Biol.* **217** 3786–96

[11] Williams N and Pearson K G 2007 Stepping of the forelegs over obstacles establishes long-lasting memories in cats *Curr. Biol.* **17** R621-623

[12] Stierle I E, Getman M and Comer C M 1994 Multisensory control of escape in the cockroach Periplaneta americana I. Initial evidence from patterns of wind-evoked behavior *J. Comp. Physiol. A* **174** 1–11

[13] Goodman B A 2007 Divergent morphologies , performance , and escape behaviour in two tropical rock-using lizards ( Reptilia : Scincidae ) 85–98

[14] Murphy R R, Tadokoro S, Nardi D, Jacoff A, Fiorini P, Choset H and Erkmen A M 2008 Search and Rescue Robotics *Handbook of Robotics* (Springer) pp 1151–73

[15] Pratt G A 2014 Robot to the rescue *Bull. At. Sci.* **70** 63–9

[16] Blaesing B and Cruse H 2004 Stick insect locomotion in a complex environment: climbing over large gaps *J. Exp. Biol.* **207** 1273–86

[17] Theunissen L M, Vikram S and Dürr V 2014 Spatial coordination of foot contacts in unrestrained climbing insects. *J. Exp. Biol.* **217** 3242–53

[18] Theunissen L M and Durr V 2013 Insects use two distinct classes of steps during unrestrained locomotion *PLoS One* **8** 1–18

[19] Harley C M, English B a and Ritzmann R E 2009 Characterization of obstacle negotiation behaviors in the cockroach, Blaberus discoidalis. *J. Exp. Biol.* **212** 1463–76

[20] Daley M A, Usherwood J R, And G F and Biewener A A 2006 Running over rough terrain: guinea fowl maintain dynamic stability despite a large unexpected change in substrate height *J. Exp. Biol.* **209**

[21] Birn-Jeffery a. V. and Daley M 2012 Birds achieve high robustness in uneven terrain through active control of landing conditions *J. Exp. Biol.* **215** 2117–27

[22] Morrey J M, Lambrecht B, Horchler a. D, Ritzmann R E and Quinn R D 2003 Highly mobile and robust small quadruped robots *Proc. 2003 IEEE/RSJ Int. Conf. Intell. Robot. Syst. (IROS 2003) (Cat. No.03CH37453)* **1** 0–5

[23] Wei T E, Quinn R D and Ritzmann R E 2005 Robot designed for walking and climbing based on abstracted cockroach locomotion mechanisms *Proceedings, 2005 IEEE/ASME Int. Conf. Adv. Intell. Mechatronics.* 24–8

[24] Chou Y-C, Yu W-S, Huang K-J and Lin P-C 2012 Bio-inspired step-climbing in a hexapod robot *Bioinspir. Biomim.* **7** 36008

[25] Johnson A M and Koditschek D E 2013 Toward a vocabulary of legged leaping *Proc. - IEEE Int. Conf. Robot. Autom.* 2568–75

[26] Chou Y C, Huang K J, Yu W S and Lin P C 2015 Model-based development of leaping in a hexapod robot *IEEE Trans. Robot.* **31** 40–54

[27] Mondada F, Pettinaro G C, Guignard A, Ivo W, Floreano D, Deneubourg J and Nolfi S 2004 Swarm-Bot : a New Distributed Robotic Concept *Auton. Robots* **17** 1–40

[28] Raibert M, Blankespoor K, Nelson G and Playter R 2008 *BigDog , the Rough-Terrain Quadruped Robot* vol 41(IFAC)

[29] Boxerbaum A S, Oro J, Peterson G and Quinn R D 2008 The latest generation Whegs robot features a passive-compliant body joint *IEEE/RSJ Int. Conf. Intell. Robot. Syst.* 1636–41

[30] Brill A L, De A, Johnson A M and Koditschek D E 2015 Tail-assisted rigid and compliant legged leaping *IEEE Int. Conf. Intell. Robot. Syst.* 6304–11

[31] More H L, Hutchinson J R, Collins D F, Weber D J, Aung S K H and Donelan J M 2010 Scaling of sensorimotor control in terrestrial mammals. *Proc. Biol. Sci.* **277** 3563–8

[32] Mongeau J-M, Sponberg S N, Miller J P and Full R J 2015 Sensory processing within cockroach antenna enables rapid implementation of feedback control for high-speed running maneuvers. *J. Exp. Biol.* **218** 2344–54

[33] Full R J and Koditschek D E 1999 Templates and anchors: neuromechanical hypotheses of legged locomotion on land *J. Exp. Biol.* **2** 3–125

[34] Spagna J C, Goldman D I, Lin P-C, Koditschek D E and Full R J 2007 Distributed mechanical feedback in arthropods and robots simplifies control of rapid running on challenging terrain. *Bioinspir. Biomim.* **2** 9–18

[35] Li C, Pullin A O, Haldane D W, Lam H K, Fearing R S and Full R J 2015 Terradynamically streamlined shapes in animals and robots enhance traversability through densely cluttered terrain *Bioinspir. Biomim* **10**

[36] Okada J and Toh Y 2000 The role of antennal hair plates in object-guided tactile orientation of the cockroach (Periplaneta americana) *J. Comp. Physiol. A* **186** 849–57

[37] Gart S, Othoyoth R, Ren Z, Yan C and Li C 2017 Dynamic locomotion of insects and legged robots over large obstacles I. Body dynamics and terrain interaction reveal a template for dynamic gap traversal *Bioinspir. Biomim*

[38] Han Y, Luo Y, Bi J and Li C 2017 Body shape affects yaw and pitch motions of insects traversing complex 3-D terrains *Integr. Comp. Biol.* 168

[39] Crall J D, Gravish N, Mountcastle A M and Combes S A 2015 BEEtag : A Low-Cost , Image-Based Tracking System for the Study of Animal Behavior and Locomotion *PLoS One* **10** e0136487

[40] Daltorio K a., Tietz B R, Bender J a., Webster V a., Szczecinski N S, Branicky M S, Ritzmann R E and Quinn R D 2013 A model of exploration and goal-searching in the cockroach, Blaberus discoidalis *Adapt. Behav.* **21** 404–20

[41] Full B Y R J and Tu M S 1990 Mechanics of six-legged runners **148** 129–46

[42] Sponberg S and Full R J 2008 Neuromechanical response of musculo-skeletal structures in cockroaches during rapid running on rough terrain. *J. Exp. Biol.* **211** 433–46

[43] Ritzmann R E, Pollack A J, Archinal J, Ridgel A L and Quinn R D 2005 Descending control of body attitude in the cockroach Blaberus discoidalis and its role in incline climbing *J. Comp. Physiol. A* **191** 253–64

[44] Goldman D I, Chen T S, Dudek D M and Full R J 2006 Dynamics of rapid vertical climbing in cockroaches reveals a template. *J. Exp. Biol.* **209** 2990–3000

[45] Daley M and Biewener A A 2006 Running over rough terrain reveals limb control for intrinsic stability. *Proc. Natl. Acad. Sci. U. S. A.* **103** 15681–6

[46] Okada J and Toh Y 2006 Active tactile sensing for localization of objects by the cockroach antenna *J. Comp. Physiol. A* **192** 715–26

[47] Qian F and Goldman D 2015 Anticipatory control using substrate manipulation enables trajectory control of legged locomotion on heterogeneous granular media Zebra-tailed *Proc. SPIE* **9467** 1–12

[48] Jusufi a, Kawano D T, Libby T and Full R J 2010 Righting and turning in mid-air using appendage inertia: reptile tails, analytical models and bio-inspired robots. *Bioinspir. Biomim.* **5** 45001

[49] Libby T, Moore T Y, Chang-Siu E, Li D, Cohen D J, Jusufi A and Full R J 2012 Tail-assisted pitch control in lizards, robots and dinosaurs *Nature* **481** 181–4

[50] Casarez C S and Fearing R S 2016 Step Climbing Cooperation Primitives for Legged Robots with a Reversible Connection *IEEE Int. Conf. Robot. Autom.* 3791–8

[51] Bibuli M, Caccia M and Lapierre L 2008 Biologically Inspired Climbing with a Hexapedal Robot *J. F. Robot.* **25** 223–42

[52] Gorb S N 2002 Structural Design and Biomechanics of Friction-Based Releasable Attachment Devices in Insects *Integr. Comp. Biol.* **42** 1127–39

[53] Daltorio K A, Wei T E, Horchler A D, Southard L, Wile G D, Quinn R D, Gorb S N and Ritzmann R E 2009 Mini-Whegs TM Climbs Attachment Mechanisms *Int. J. Rob. Res.* **28** 285–302

[54] Cowan N J and Full R J 2006 Task-level control of rapid wall following in the American

cockroach *J. Exp. Biol.* **209** 3043–3043

[55] Daltorio K A, Mirletz B T, Sterenstein A, Cheng J C, Watson A, Kesavan M, Bender J A, Ritzmann R E and Quinn R D 2014 How Cockroaches Employ Wall-Following for Exploration 72–83

[56] Ting L H, Blickhan R and Full R J 1994 Dynamic and Static Stability in Hexapedal Runners *J. Exp. Biol.* **197** 251–69

[57] Ritzmann J T W R E 1998 Leg kinematics and muscle activity during treadmill running in the cockroach , Blaberus discoidalis : II . Fast running *J. Comp. Physiol. A* **182** 23–33

[58] Ritzmann J T W R E 1998 Leg kinematics and muscle activity during treadmill running in the cockroach , Blaberus discoidalis : I . Slow running *J. Comp. Physiol. A* **182** 11–22

[59] Tryba A K and Ritzmann R O Y E 2000 Multi-Joint Coordination During Walking and Foothold Searching in the Blaberus Cockroach . II . Extensor Motor Neuron Pattern *J. Neurophysiol.* **83** 3337–50

[60] Tryba A K and Ritzmann R O Y E 2000 Multi-Joint Coordination During Walking and Foothold Searching in the Blaberus Cockroach . I . Kinematics and Electromyograms *J. Neurophysiol.* **83** 3323–36

[61] Ridgel A L and Ritzmann R E 2005 Effects of neck and circumoesophageal connective lesions on posture and locomotion in the cockroach *J. Comp. Physiol. A* **191** 559–73

[62] Leonard J J and Durrant-Whyte H F 1991 Simultaneous map building and localization for an autonomous mobile robot *Intell. Robot. Syst.* **3** 1442–7

[63] Thrun S, Burgard W and Fox D 2000 A real-time algorithm for mobile robot mapping with applications to\nmulti-robot and 3D mapping *IEEE Int. Conf. Robot. Autom.* **1** 321–8

[64] Bell W J, Roth L M and Nalepa C A 2007 *Cockroaches. Ecology, Behaviour and Natural History* (Baltimore, MD, USA: The Johns Hopkins University Press)

[65] Michalková V and Pekár S 2009 How glyphosate altered the behaviour of agrobiont spiders ( Araneae : Lycosidae ) and beetles ( Coleoptera : Carabidae ) *Biol. Control* **51** 444–9

[66] Griffiths E, Wratten S and Vickerman G 1985 Foraging by the carabid Agonum dorsde in the field *Ecol. Entomol.* **10** 181–9

# Supplementary Figures

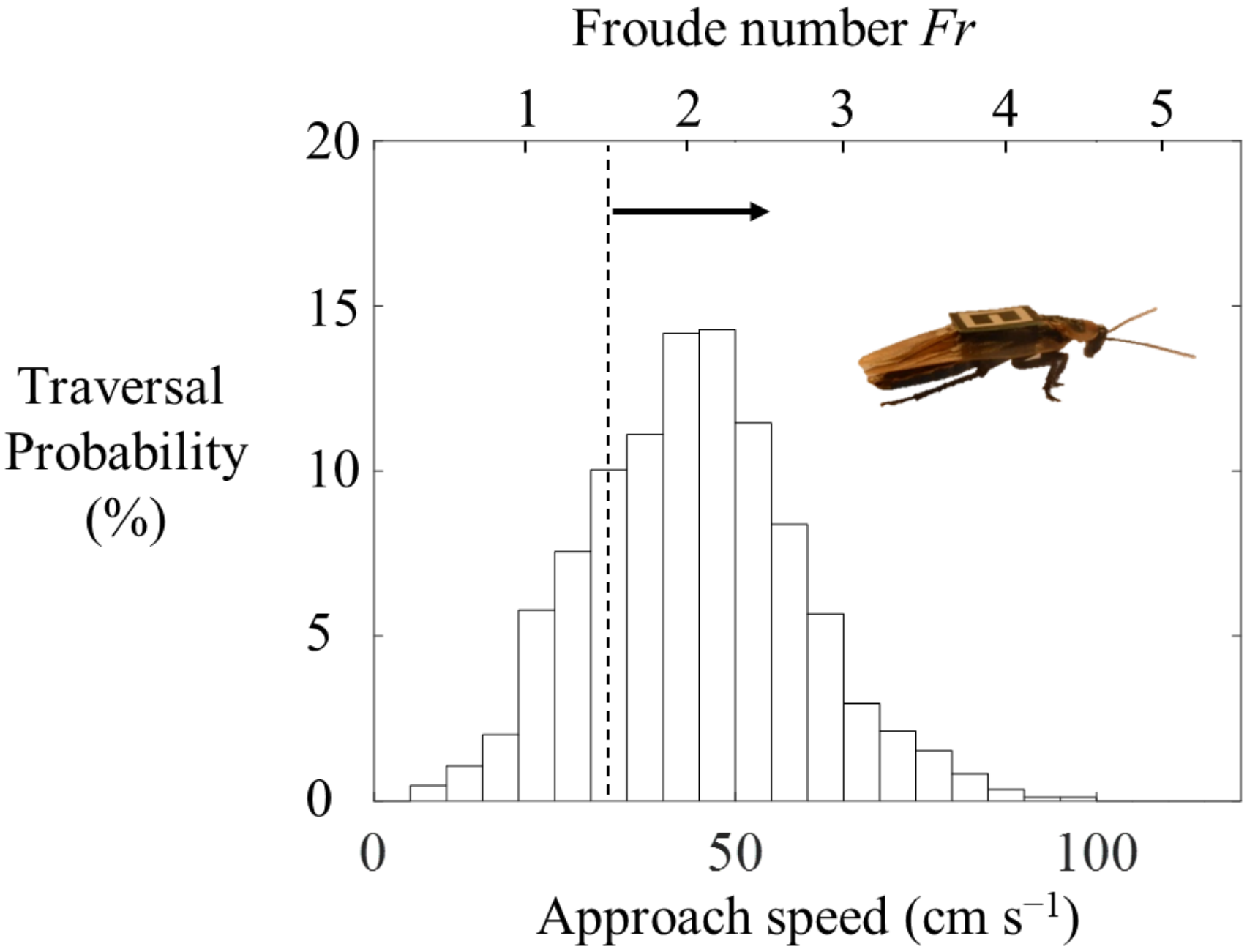

**Figure S1.** Histogram of approach speeds for animal bump experiments. Vertical dashed line represents an approach speed of 30 cm $s^{-1}$, $Fr$ = 1.5, above which value the animals are unlikely to react obstacle encounter (see section 3.3).

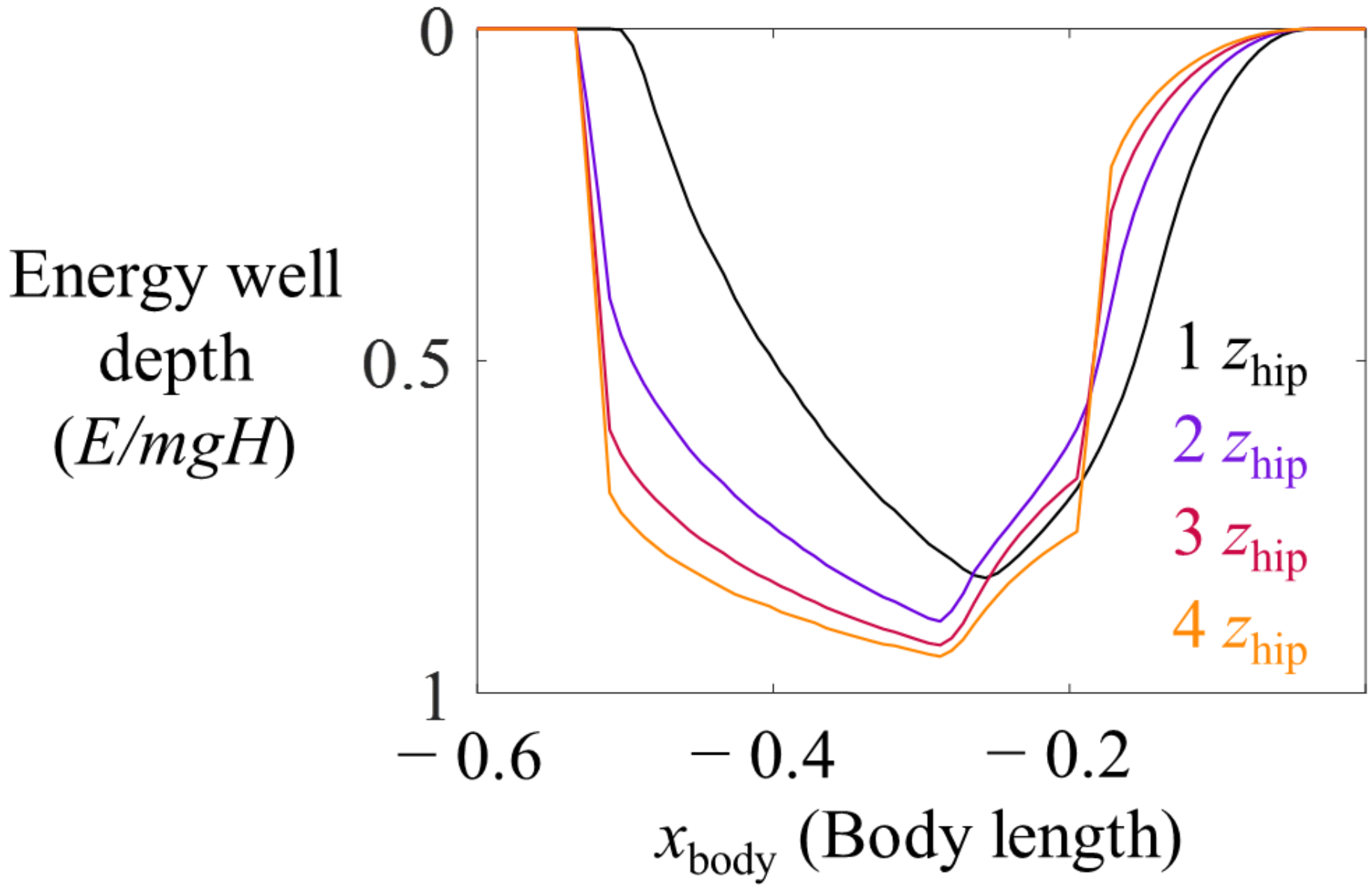

**Figure S2.** Energy well depth as a function of forward position of the body for each bump height. The energy well depth for each forward position of the body was calculated by taking the difference between the potential energy of a body with zero body pitch and body yaw (figure 9(b, d, f), black circle) and the minimal potential energy on the landscape (for example, figure 9(b), 0° body pitch).

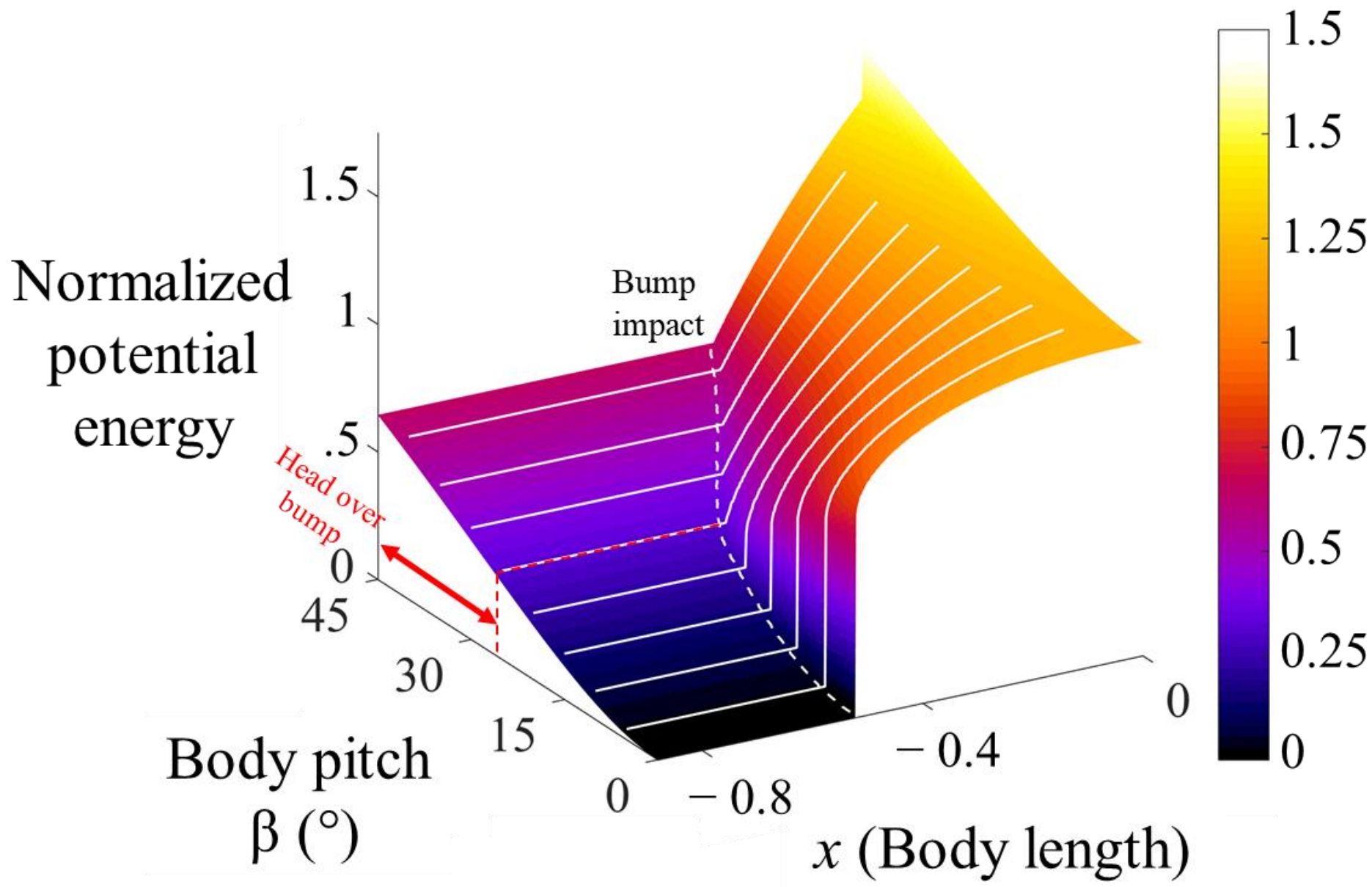

**Figure S3.** Locomotion energy landscape as a function of body pitch and forward position of the body (shown for the 4 hip height bump as an example). Potential energy is normalized by body mass and bump height. The white arrows show the potential energy of the body at representative initial body pitch angles as it translates towards the bump. The white dashed curve on the surface shows where the body contacts the bump. The red arrow shows the range of body pitch where the head reaches over the near top corner of the bump.

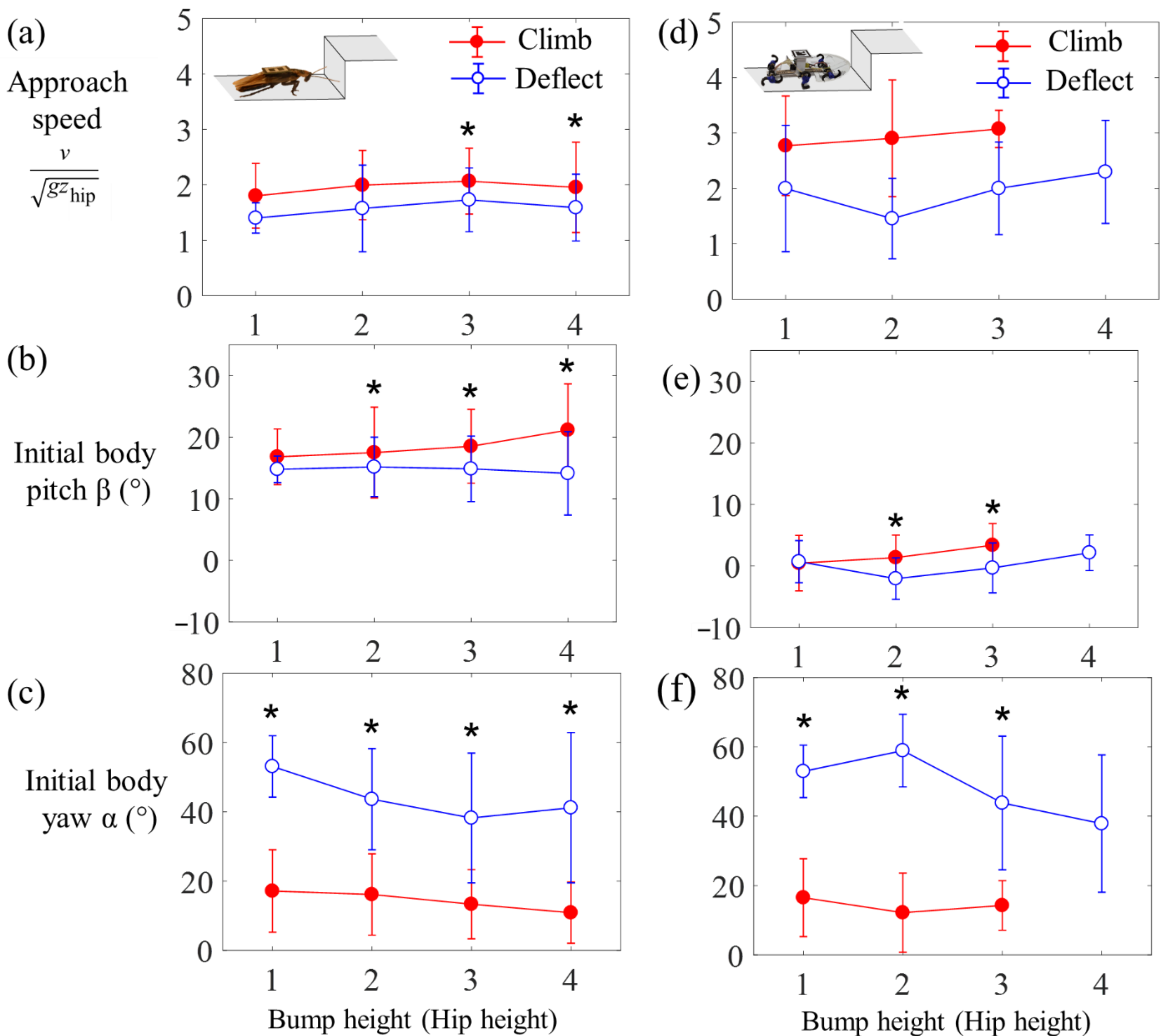

**Figure S4.** Approach kinematics as function of bump height for the animal (a, b, c) and robot (d, e, f) experiments. (a, d) Initial body pitch as function of bump height. (b, e) Approach speed as function of bump height. (c, f) Initial body yaw as function of bump height. Red filled circles represent climbing traversals and blue open circles represent deflections. Error bars represent ± 1 s.d. Asterisks in indicate a statistically significance difference between the cases of climbing and deflection ($P < 0.05$, multiple logistic regression).

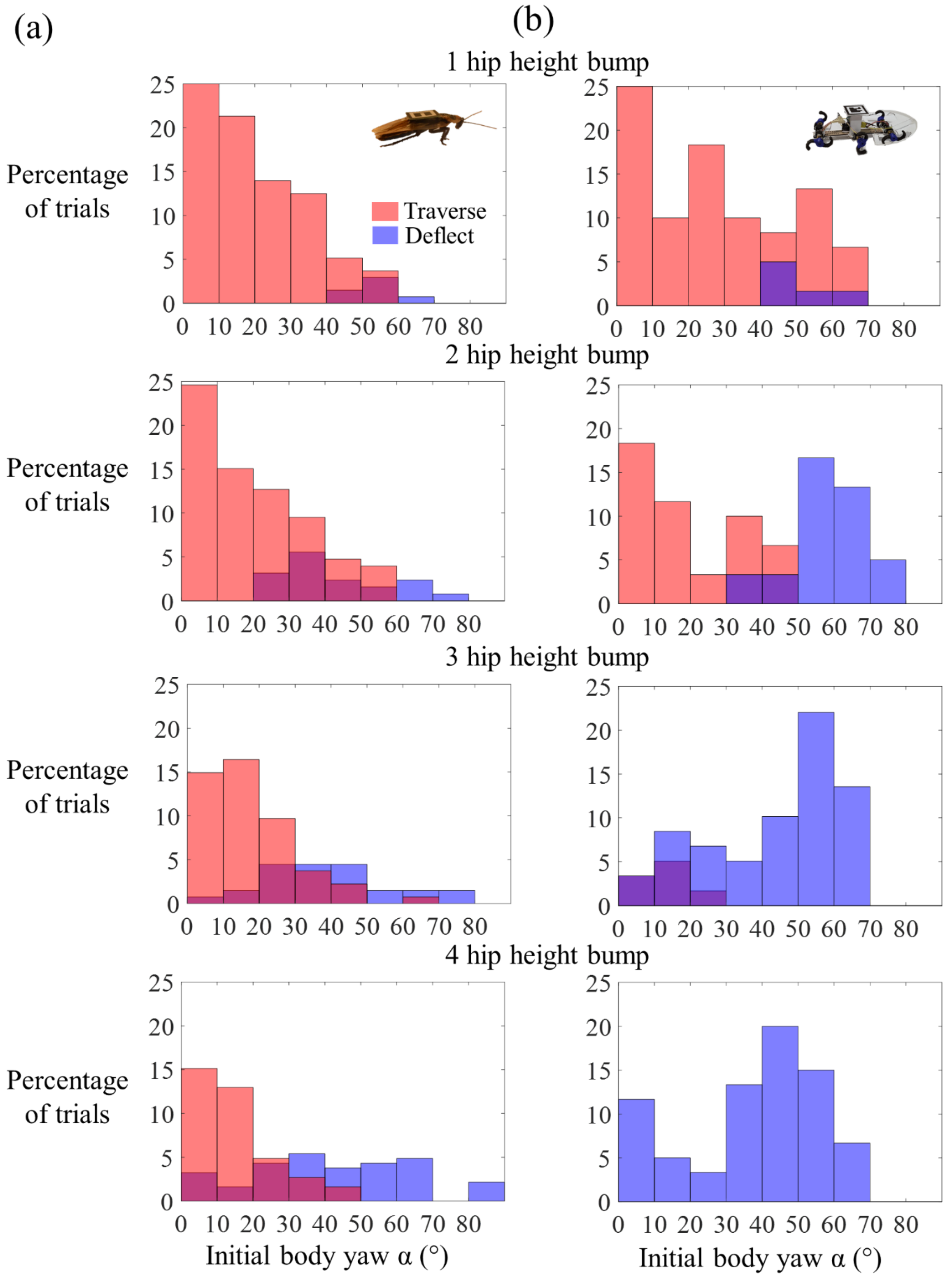

**Figure S5.** Histogram of initial body yaw for trials that traverse the bump and trials that deflect and fail to traverse for animals (a) and robot (b).

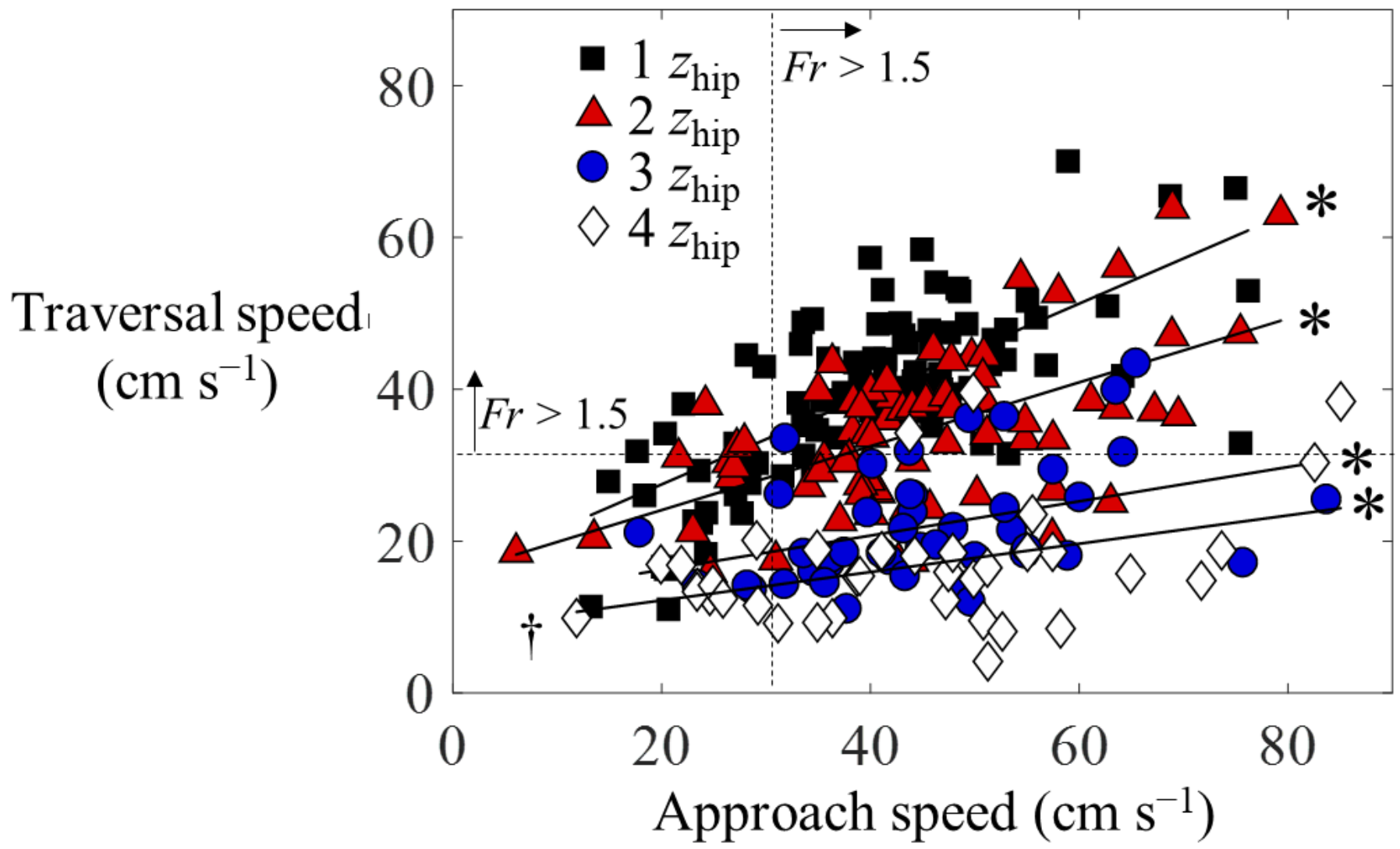

**Figure S6.** Traversal speed as a function of approach speed. The dashed lines show the animal's walking-to-running transition speed. Asterisks represent significant non-zero slopes (linear least squares regression). †Shows the of approach speed for previous studies of cockroach bump traversal [1], [2].

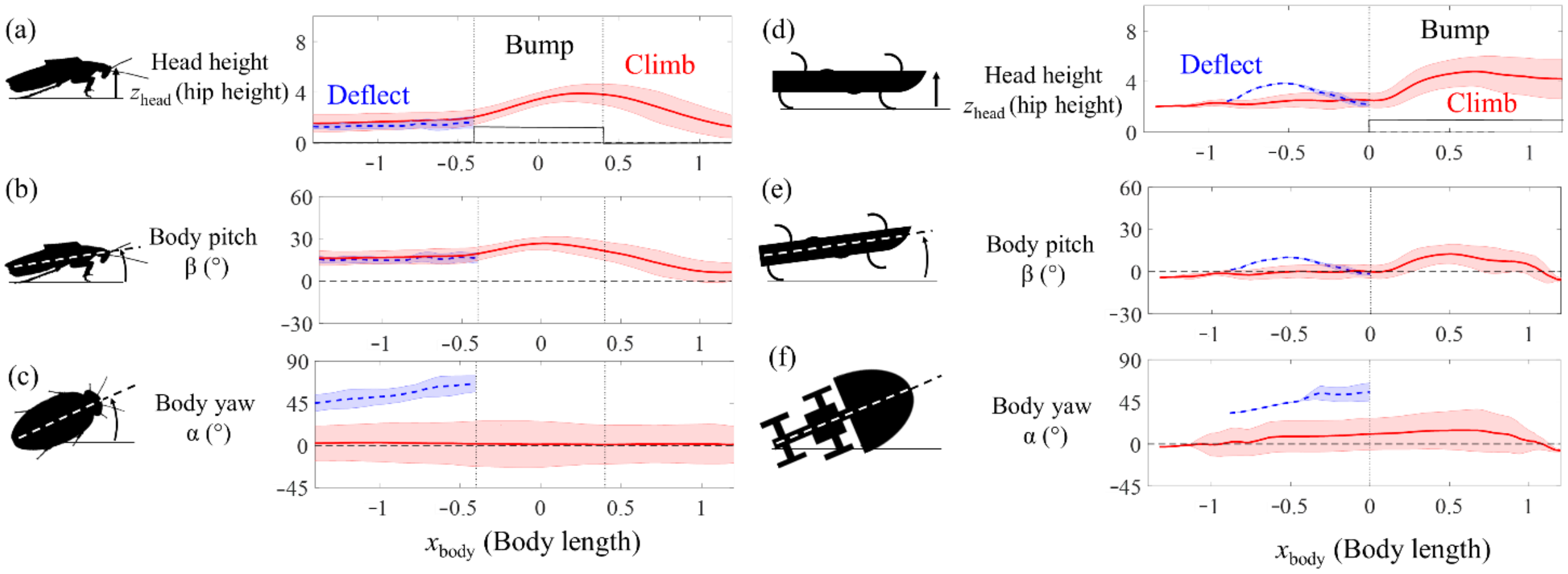

**Figure S7.** Dynamic locomotion of the animal (a, b, c, d) and robot (e, f, g, h) over a 1 hip height bump obstacle. (a, e) Head height as a function of forward of the head. (b, f) Body pitch as a function of forward of the head. (c, g) Body roll as a function of forward of the head. (d, h) Body yaw as a function of forward of the head. Solid red and blue dashed curves and shaded areas represent means ± 1 s.d. for the cases of climbing and deflection. For simplicity, only movement perpendicular to the bump (within the *x-z* plane) is shown.

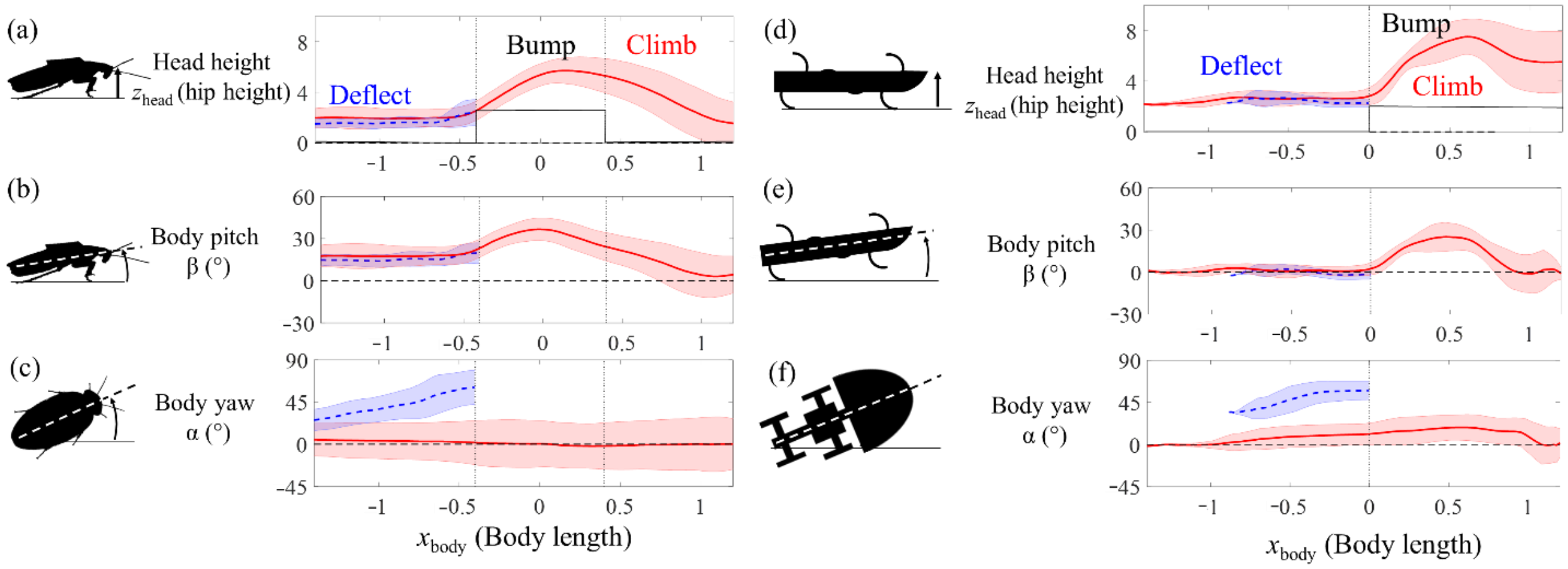

**Figure S8.** Dynamic locomotion of the animal (a, b, c, d) and robot (e, f, g, h) over a 2 hip height bump obstacle. (a, e) Head height as a function of forward of the head.* (b, f) Body pitch as a function of forward of the head. (c, g) Body roll as a function of forward of the head. (d, h) Body yaw as a function of forward of the head. Solid red and blue dashed curves and shaded areas represent means ± 1 s.d. for the cases of climbing and deflection. For simplicity, only movement perpendicular to the bump (within the *x*-*z* plane) is shown.

* The slight "penetration" of the head into the bump in (a, d) was an artifact due to bending of the neck joint of the animal or deformation of the robot anterior during collision).

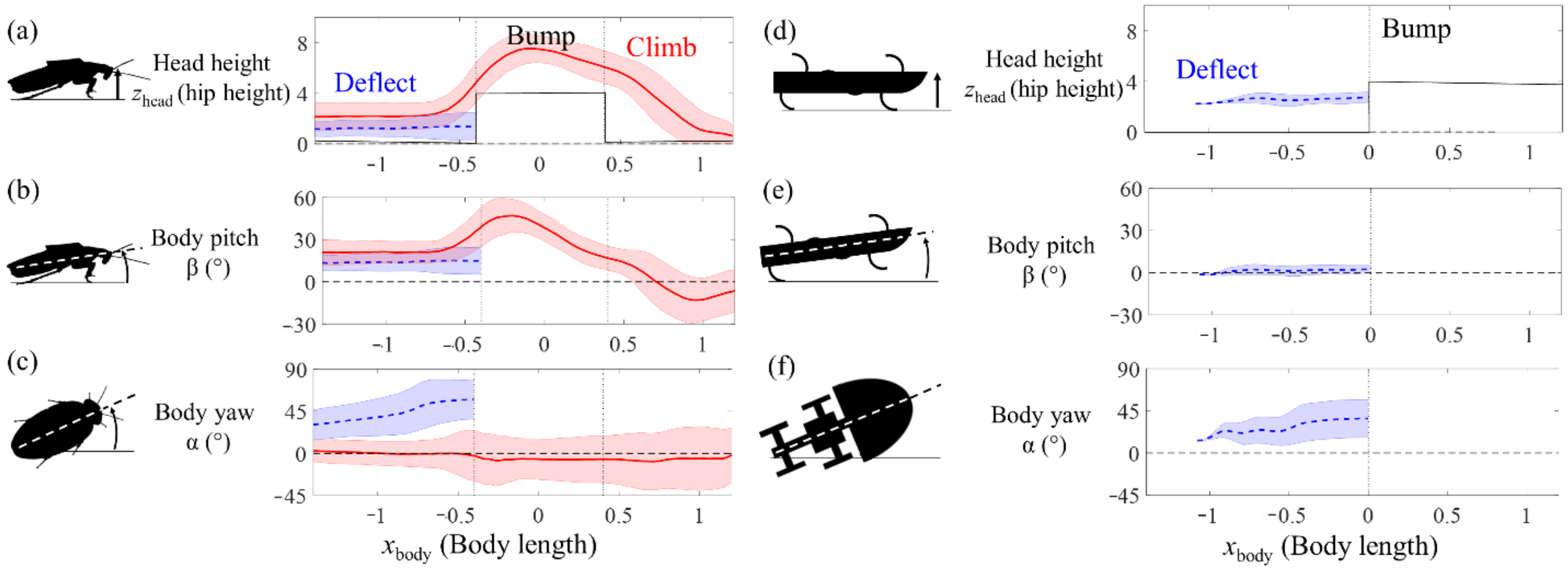

**Figure S9.** Dynamic locomotion of the animal (a, b, c, d) and robot (e, f, g, h) over a 4 hip height bump obstacle. (a, e) Head height as a function of forward of the head.* (b, f) Body pitch as a function of forward of the head. (c, g) Body roll as a function of forward of the head. (d, h) Body yaw as a function of forward of the head. Solid red and blue dashed curves and shaded areas represent means ± 1 s.d. for the cases of climbing and deflection. For simplicity, only movement perpendicular to the bump (within the *x*-*z* plane) is shown.

* The slight "penetration" of the head into the bump in (a, d) was an artifact due to bending of the neck joint of the animal or deformation of the robot anterior during collision).

Supplementary Video 1

https://www.youtube.com/watch?v=7ZfMls2l5O0

Supplementary Video 2

https://www.youtube.com/watch?v=_I5vM6PgSbY

Supplementary Video 3

https://www.youtube.com/watch?v=v1vXGOw5I_Q

Supplementary Video 4

https://www.youtube.com/watch?v=ouPlzMfjN4k